\documentclass[preprintnumbers,superscriptaddress,showkeys,showpacs,byrevtex]{revtex4}
\usepackage{amsmath,amsfonts,amssymb,amscd,amsxtra,amsthm}
\usepackage{graphicx}
\usepackage{bm}
\usepackage{graphics}
\usepackage{amsmath}
\usepackage{epsfig}
\usepackage{epsf}
\usepackage{epstopdf}
\usepackage{multirow}
\begin{document}
\preprint{CYCU-HEP-13-03}
\preprint{KIAS-P13014}
\title{The quark-jet contribution to the fragmentation functions
\\ for the pion and kaon with the nonlocal interactions}
\author{Dong-Jing Yang}
\email[E-mail: ]{djyang@std.ntnu.edu.tw}
\affiliation{Department of Physics, National Taiwan Normal University, Taipei 11677, Taiwan}
\author{Fu-Jiun Jiang}
\email[E-mail: ]{fjjiang@ntnu.edu.tw}
\affiliation{Department of Physics, National Taiwan Normal University, Taipei 11677, Taiwan}
\author{Chung-Wen Kao}
\email[E-mail: ]{cwkao@cycu.edu.tw}
\affiliation{Department of Physics, Chung-Yuan Christian University, Chung-Li 32023, Taiwan}
\author{Seung-il Nam}
\email[E-mail: ]{sinam@kias.re.kr}
\affiliation{School of Physics, Korea Institute for Advanced Study (KIAS), Seoul 130-722, Korea}
\date{\today}
\begin{abstract}
We investigate the unpolarized pion and kaon fragmentation functions using the nonlocal chiral-quark model.
In this model the
interactions between the quarks and pseudoscalar mesons is manifested nonlocally. In addition, the explicit flavor SU(3) symmetry
breaking effect is taken into account in terms of the current quark masses.
The results of our model are evaluated to higher $Q^2$ value $Q^2=4\,\mathrm{GeV}^2$ by the DGLAP evolution.
Then we compare them with the empirical parametrizations.
We find that our results are in relatively good agreement with the empirical parametrizations
and the other theoretical estimations.
\end{abstract}
\maketitle
\section{Introduction}
Fragmentation functions are important ingredient for understanding the structure of the hadrons, because they play a crucial role in analyzing the
processes involving hadrons. For example, one needs the unpolarized fragmentation functions to analyze the semi-inclusive
processes in the electron-positron scattering, deep-inelastic proton-proton scattering, and
so on~\cite{Collins:1992kk,Mulders:1995dh,Boer:1997nt,Anselmino:1994tv, Anselmino:2008jk,Christova:2006qs,Anselmino:2007fs,Bacchetta:2006tn,
Efremov:2006qm,Collins:2005ie,Ji:2004wu}.
Furthermore, to extract the chiral-odd transversity parton distribution of the nucleon one needs more complicated fragmentation functions, such as the
polarized dihadron fragmentation functions and the Collins fragmentation functions. Because of their fundamental importance,
those functions have been studied intensively for decades but still not fully understood yet.

The unpolarized fragmentation function $D^H_q(z)$ represents the probability for a quark $q$ to emit a hadron $H$ with the light-cone momentum fraction
$z$. It can be written with the light-cone coordinate as follows~\cite{Bacchetta:2002tk,Amrath:2005gv}:
\begin{equation}
\label{eq:FRAG}
D^H_q(z,\bm{k}^2_T,\mu)=\frac{1}{4z}\int dk^+\mathrm{Tr}
\left[\Delta(k,p,\mu)\gamma^- \right]|_{zk^-=p^-}.
\end{equation}
Here, $k_{\pm}$=$(k_{0}\pm k_{3})/\sqrt{2}$ and the correlation $\Delta(k,p,\mu)$ is defined as
\begin{equation}
\Delta(k,p,\mu)=\sum_X\int\frac{d^4\xi}{(2\pi)^4}e^{+ik\cdot\xi}
\langle0|\psi(\xi)|H,X\rangle\langle H,X|\psi(0)|0\rangle,
\label{eq:COR1}
\end{equation}
where $k$, $p$ indicate the four-momenta for the initial quark and fragmented hadron, respectively. Furthermore, $z$ is the longitudinal light-cone
momentum fraction possessed by the hadron and $\mu$ denotes a renormalization scale at which the fragmentation process is computed.
All the calculations are carried out in the frame where the $z$-axis is chosen to be the direction of $\bm{k}$. Consequently the transverse momentum of
the initial quark $\bm{k}_{\perp}$ is zero in this frame. On the other hand, $\bm{k}_{T}=\bm{k}-[(\bm{k}\cdot\bm{p})/|\bm{p}|^2]\,\bm{p}$, defined as the
transverse momentum of the initial quark with respect to the direction of the momentum of the produced hadron, is nonzero. The integrated fragmentation
function satisfies the momentum sum rule:
\begin{equation}
\label{eq:SUM}
\int^1_0\sum_{H}zD^H_{q}(z,\mu)\,dz=1,
\end{equation}
where $H$ represents all the fragmented hadrons. Eq.~(\ref{eq:SUM}) means that all of the momentum of the initial quark $q$ is transferred
into the momenta of the fragmented hadrons. Empirically, information of $D^H_{q}(z)$  has to be extracted from the available high-energy lepton-scattering
data by global analysis with appropriate parameterizations satisfying
certain constraints~\cite{Kretzer:2000yf,Conway:1989fs,deFlorian:2007aj,Sutton:1991ay,Hirai:2007cx,Kniehl:2000fe}.

From theoretical points of view,
it is impossible to study fragmentation functions directly by lattice QCD because they are defined
in Minkowski space. Nevertheless, there have been numerous works for the fragmentation
functions based on the effective QCD models so far.
In Ref.~\cite{Ito:2009zc}, the Nambu--Jona-Lasinio (NJL) model has been used to calculate the
fragmentation functions. Monte-Carlo (MC) simulations with supersymmetric QCD were also carried out to obtain the fragmentation function up to a very
high energy in the center-of-mass frame $\sqrt{s}$~\cite{Aloisio:2003xj}. The Collins fragmentation functions, which play an essential role in the
transverse-spin physics, have been studied as well in the quark-pseudoscalar (PS) meson coupling
model~\cite{Bacchetta:2002tk,Amrath:2005gv,Bacchetta:2007wc}. Note that dihadron fragmentation functions have been
investigated in the same theoretical
formalism~\cite{Bacchetta:2006un}. For brevity,
we will simply call the unpolarized fragmentation functions as the fragmentation functions from now on.

In Refs.~\cite{Nam:2011hg,Nam:2012af}, we have already employed the nonlocal chiral quark model (NLChQM)
with the explicit flavor SU(3) symmetry breaking to calculate the elementary fragmentation
functions. This instanton-motivated approaches were used for computing the
quark distribution amplitudes, manifesting the nonlocal quark-pseudoscalar (PS) meson
interactions~\cite{Dorokhov:1991nj,Nam:2006sx,Nam:2006au,Praszalowicz:2001pi}. NLChQM have been applied to determine various nonperturbative
quantities and obtained the results which are in good agreement with experiments as well as lattice QCD (LQCD)
simulations~\cite{Musakhanov:1998wp,Musakhanov:2002vu,Nam:2007gf,Nam:2010pt,Dorokhov:2000gu,Dorokhov:2002iu}.
The elementary fragmentation functions
calculated in Refs.~\cite{Nam:2011hg,Nam:2012af} are the functions in Eq.(\ref{eq:COR1}) with the following approximation:
\begin{equation}
\sum_{X}|h,X\rangle \langle h,X| \approx \sum_{h}|h=q\bar{Q},\,X=Q\rangle \langle h=q\bar{Q},\,X=Q|.
\label{eFF}
\end{equation}
Here $h$ denotes the PS meson. In other words, we just calculate the one-step fragmentation process: $q\rightarrow h=(q\bar{Q})+Q$.
\begin{figure}[t]
\begin{tabular}{c}
\includegraphics[width=8.5cm]{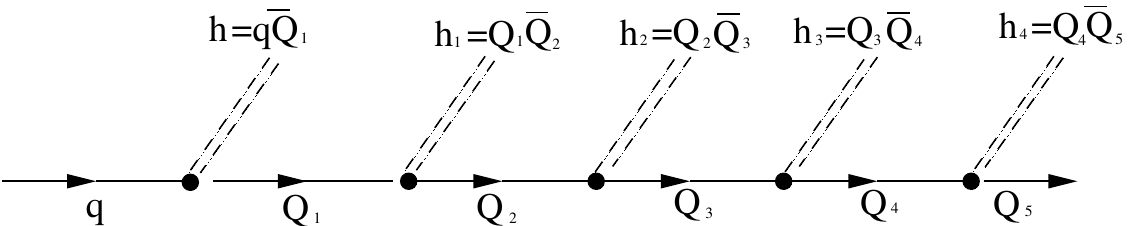}
\end{tabular}
\caption{Quark fragmentation cascade process.}
\label{cascade}
\end{figure}
In Refs.~\cite{Nam:2011hg,Nam:2011hg}, the renormalized fragmentation functions are obtained by re-scaling the elementary fragmentation functions.
After DGLAP evolution the results of the renormalized fragmentation functions agree with the empirical ones reasonably well except at small $z$.
However such renormalized fragmentation functions do not satisfy the sum rules of Eq.~(\ref{eq:SUM}). Furthermore some of the fragmentation functions , such as $u \rightarrow \pi^{-}$ and $s\rightarrow K^{+}$, are zero because their elementary fragmentation functions are
identical zero.
Hence it is necessary to include the quark-jet contribution from many-step fragmentation
processes.

In Refs.~\cite{Ito:2009zc,Matevosyan:2010hh,Matevosyan:2011ey,Matevosyan:2011vj}, the NJL model has been applied
for the fragmentation functions including
the quark-jets and resonances. The momentum sum rules is satisfied automatically according to their approach.
It turned out that the quark-jet contributions provide
considerable contributions to the various fragmentation functions at the small $z$ region. The approach in
\cite{Ito:2009zc,Matevosyan:2010hh,Matevosyan:2011ey,Matevosyan:2011vj} is actually applicable for any effective model. With this method,
one can generate the fragmentation functions $D^{h}_{q}(z)$ by solving the coupled differential-integral equations where the elementary fragmentation
functions $d^{h}_{q}(z)$ appear in the kernel. In this article, we will extend our previous works in \cite{Nam:2011hg,Nam:2012af} by including the quark-jet contribution of the fragmentation functions using the method developed in
\cite{Ito:2009zc,Matevosyan:2010hh,Matevosyan:2011ey,Matevosyan:2011vj}. We will evolved our results to higher $Q^2$ values.
Then we will compare our results with the empirical parameterizations and the results from the NJL-jet model.

The present work is organized as follows: In Section II, we briefly sketch our results of the elementary fragmentation functions based on NLChQM. In
Section III, we explain the method of calculating the quark-jet contribution to the fragmentation functions and apply this method to our model.
In Section IV, we present and discuss our numerical results of fragmentation functions at $Q^2=4\,\mathrm{GeV}^2$. The final Section is devoted for the conclusions and future perspectives.
\section{Elementary fragmentation functions in NLChQM}
In this section, we briefly explain how to derive the elementary fragmentation functions in NLChQM.
This model is motivated from the dilute
instanton liquid model(DILM)~\cite{Diakonov:1985eg,Shuryak:1981ff,Diakonov:1983hh,Diakonov:2002fq,Schafer:1996wv} in which
the nonperturbative QCD effects
are from the nontrivial quark-instanton interactions in the dilute instanton ensemble.
However DILM is defined in Euclidean space since the
(anti)instantons are well defined there as the tunneling between the infinitely degenerate QCD vacua.
However, there are still some works
which generalize DILM to calculate some physical quantities such as the light-cone wave function which are properly defined only in Minkowski
space ~\cite{Dorokhov:1991nj,Nam:2006sx,Nam:2006au,Praszalowicz:2001pi}. Following those studies, we adopt the effective chiral action (EChA) from
NLChQM in Minkowski space as follows:
\begin{equation}
\label{eq:ECA}
\mathcal{S}_\mathrm{eff}[m_q,h]=-i\mathrm{Sp}\ln\left[i\rlap{/}{\partial}
-\hat{m}_f-\sqrt{M(\loarrow{\partial}^2)}U^{\gamma_5}\sqrt{M(\roarrow{\partial}^2)}\right],
\end{equation}
where $\mathrm{Sp}$ and $\hat{m}_q$ denote the functional trace $\mathrm{Tr}\int d^4x \langle x|\cdots|x\rangle$ over all the relevant spin spaces and
SU(3) current-quark mass matrix, $\mathrm{diag}(m_u,m_d,m_s)$, respectively.
Here we choose the following values: $(m_u,m_d,m_s)=(5,5,150)$ MeV.
As mentioned in Refs.~\cite{Nam:2011hg,Nam:2012af}, to derive EChA Eq.~(\ref{eq:ECA}) we simply replace the Euclidean metric for the (anti)instanton
effective chiral action by the one of Minkowski space.
The interactions between the quarks and the nonperturbative QCD vacuum generate the momentum-dependent effective quark mass
which can be written in a simple $n$-pole type form factor as follows~\cite{Dorokhov:1991nj,Nam:2006sx,Nam:2006au,Praszalowicz:2001pi}:
\begin{equation}
\label{eq:MDM}
M(\partial^2)=M_0\left[\frac{n\Lambda^2}{n\Lambda^2-\partial^2+i\epsilon} \right]^{n},
\end{equation}
where $n$ indicates a positive integer number. We will choose $n=2$ as in the instanton model~\cite{Nam:2006sx,Nam:2006au}.
In Eq.~(\ref{eq:MDM}) $\Lambda\approx\mu$ stands for the model renormalization scale. It is related to the average (anti)instanton size
$\bar{\rho}$ in principle and takes the value $\Lambda\approx600$ MeV~\cite{Diakonov:2002fq}. The
nonlinear PS-meson field, i.e. $U^{\gamma_5}$ takes a simple form (i.e.~Ref.~\cite{Diakonov:2002fq}) with the normalization chosen
to be consistent with the definition of the fragmentation function in Eq.~(\ref{eq:FRAG})~\cite{Bacchetta:2002tk}:
\begin{equation}
\label{eq:CHIRALFIELD}
U^{\gamma_5}(h)=
\exp\left[\frac{i\gamma_{5}(\bm{\lambda}\cdot\bm{h})}{2F_h}\right]
=1+\frac{i\gamma_{5}(\bm{\lambda}\cdot\bm{h})}{2F_h}
-\frac{(\bm{\lambda}\cdot \bm{h})^{2}}{8F^{2}_h}+\cdots,
\end{equation}
where $F_ h$ and $\lambda^a$ stand for the weak-decay constant for the PS meson $h$ and the Gell-Mann matrix.
The flavor SU(3) octet PS-meson fields are given as:
\begin{equation}
\label{eq:PHI}
\bm{\lambda}\cdot\bm{h}=\sqrt{2}\left(
\begin{array}{ccc}
\frac{1}{\sqrt{2}}\pi^{0}+\frac{1}{\sqrt{6}}\eta&\pi^{+}&K^+\\
\pi^-&-\pi^{0}+\frac{1}{\sqrt{6}}\eta&K^0\\
K^-&\overline{{K}^0}&-\frac{2}{\sqrt{6}}\eta\\
\end{array} \right).
\end{equation}
By expanding the nonlinear PS-meson field from EChA in Eq.~(\ref{eq:ECA}), one derives the following effective interaction Lagrangian
density {in the coordinate space} for the nonlocal quark-quark-PS meson vertex:
\begin{equation}
\label{eq:LAG}
\mathcal{L}_{qQh}=\frac{i}{2F_h}\bar{Q}\sqrt{M(\loarrow{\partial}^2)}
\gamma_5(\bm{\lambda}\cdot \bm{h})\sqrt{M(\roarrow{\partial}^2)}q.
\end{equation}
As a result, we reach a concise expression for the elementary fragmentation function $q(k)\to h(p)+Q(r)$ from NLChQM:
\begin{equation}
\label{eq:FRAGDEF}
d^{h}_{q}(z,\bm{k}^2_T,\mu)=\frac{\mathcal{C}^{h}_{q}}{8\pi^3z(1-z)}\frac{M_kM_{r}}{2F^2_h}
\left[\frac{z(k^2-\bar{M}^2_q)+(k^2+\bar{M}^2_q-2\bar{M}_q\bar{M}_{Q}-2k\cdot p) }{(k^2-\bar{M}^2_q)^2}\right],
\end{equation}
where $\mathcal{C}^h_{q}$ indicates the flavor factor for the corresponding fragmentation processes and is given in Table~\ref{TABLE0} in the Appendix.
Notice the flavor for the initial quark $q$ in Eq.~(\ref{eq:FRAGDEF}) is written explicitly. Furthermore, the momentum dependent effective quark mass
reads:
\begin{equation}
\label{eq:MDQM}
M_\ell=M_0\left[\frac{2\Lambda^2}{2\Lambda^2-\ell^2+i\epsilon} \right]^{2}.
\end{equation}
Here we have used the notation: $\bar{M}_q\equiv m_q+M_0$. The value of $M_0$ can be fixed self-consistently within the instanton
model~\cite{Diakonov:1985eg,Shuryak:1981ff,Diakonov:1983hh,Schafer:1996wv,Musakhanov:1998wp,Musakhanov:2002vu,Diakonov:2002fq,Nam:2007gf,Nam:2010pt}
with the phenomenological (anti)instanton parameters
$\bar{\rho}\approx1/3$ fm and $\bar{R}\approx1$ fm. This will lead to $M_0\approx350$ MeV.
The masses for the pion and kaon are chosen to be $m_{\pi,K}=(140,495)$
MeV throughout the present work. Taking all the considerations into account, one arrives at a concise expression for the elementary fragmentation functions:
\begin{equation}
\label{eq:DDDDD}
d^{h}_{q}(z,\bm{k}^2_T,\mu)=\frac{\mathcal{C}^h_{q}}
{8\pi^3}\frac{M_kM_{r}}{2F^2_h}
\frac{z\left[z^2\bm{k}^2_T+[(z-1)\bar{M}_q+\bar{M}_{Q}]^2\right]}
{[z^2\bm{k}^2_T+z(z-1)\bar{M}^2_q+z\bar{M}^2_{Q}+(1-z)m^2_h]^2},
\end{equation}
where $M_k$ and $M_{r}$ are the momentum-dependent quark masses manifesting the nonlocal quark-PS meson interactions:
\begin{equation}
\label{eq:MASS}
M_k=\frac{M_0[2\Lambda^2z(1-z)]^2}
{[z^2\bm{k}^2_T+z(z-1)(2\Lambda^2-\delta^2)+z\bar{M}^2_{Q}+(1-z)m^2_h]^2},
\,\,\,\,
M_{r}=\frac{M_0(2\Lambda^2)^2}{(2\Lambda^2-\bar{M}^2_{Q})^2}.
\end{equation}
As in Ref.~\cite{Nam:2011hg}, a free and finite-valued parameter $\delta$ has been
introduced in the denominator to avoid the unphysical singularities.
Notice the singularities arise in the vicinity of $(z,\bm{k}_T)=0$ due the present parametrization of the effective quark mass as in Eq.~(\ref{eq:MDM}).
At the renormalization scale in our model, the elementary fragmentation function can be evaluated further by integrating Eq.~(\ref{eq:DDDDD})
over $k_T$:
\begin{equation}
\label{eq:FRAGINT}
d^{h}_{q}(z,\mu)=2\pi z^2\int^\infty_0 d^{h}_{q}(z,\bm{k}^2_T,\mu)
\,\bm{k}_T\,d\bm{k}_T.
\end{equation}

\section{Fragmentation functions with the quark-jet contribution}
\begin{figure}[t]
\begin{tabular}{cc}
\includegraphics[width=8.5cm]{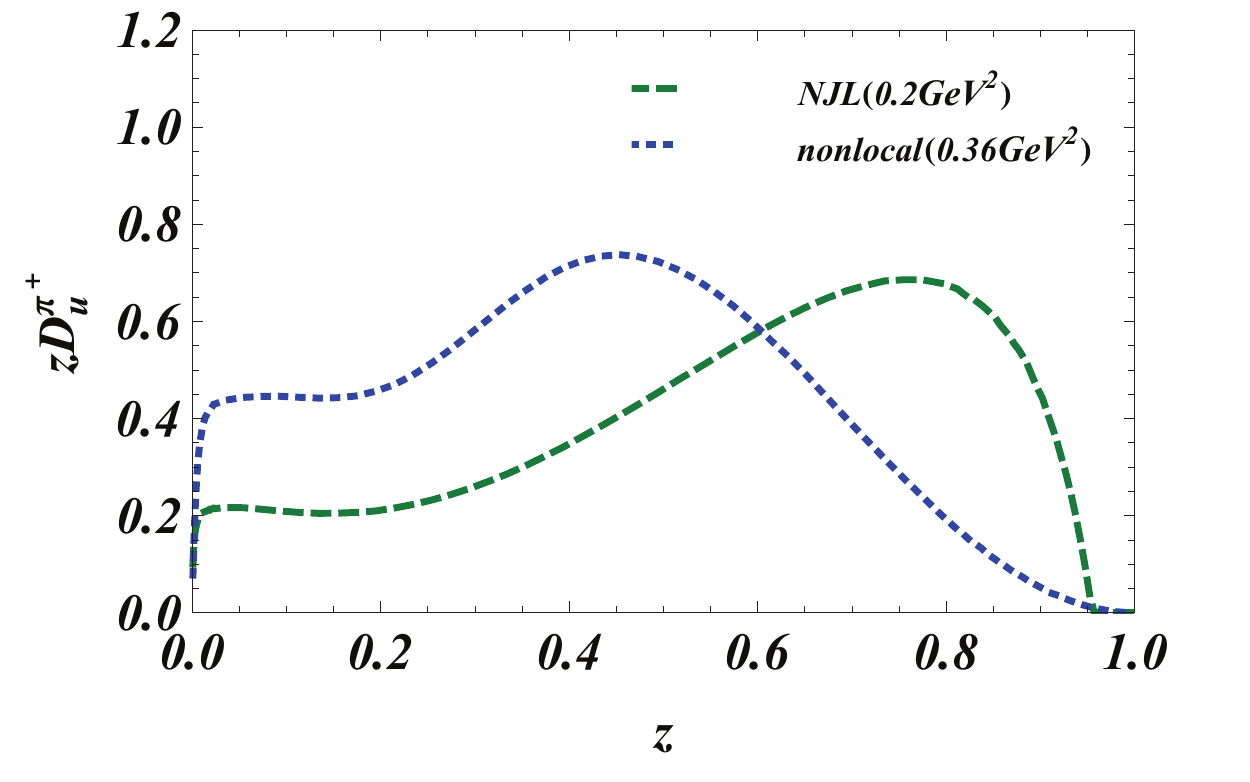}
\includegraphics[width=8.5cm]{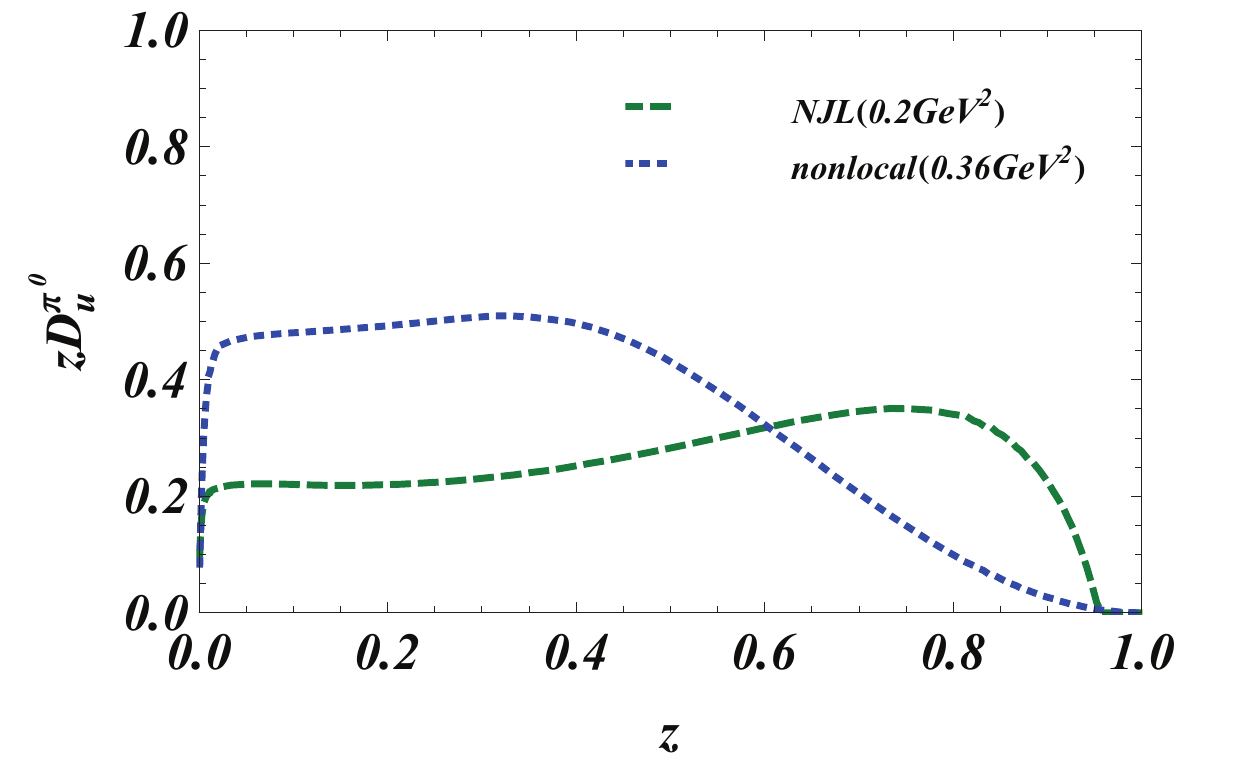}
\end{tabular}
\begin{tabular}{cc}
\includegraphics[width=8.5cm]{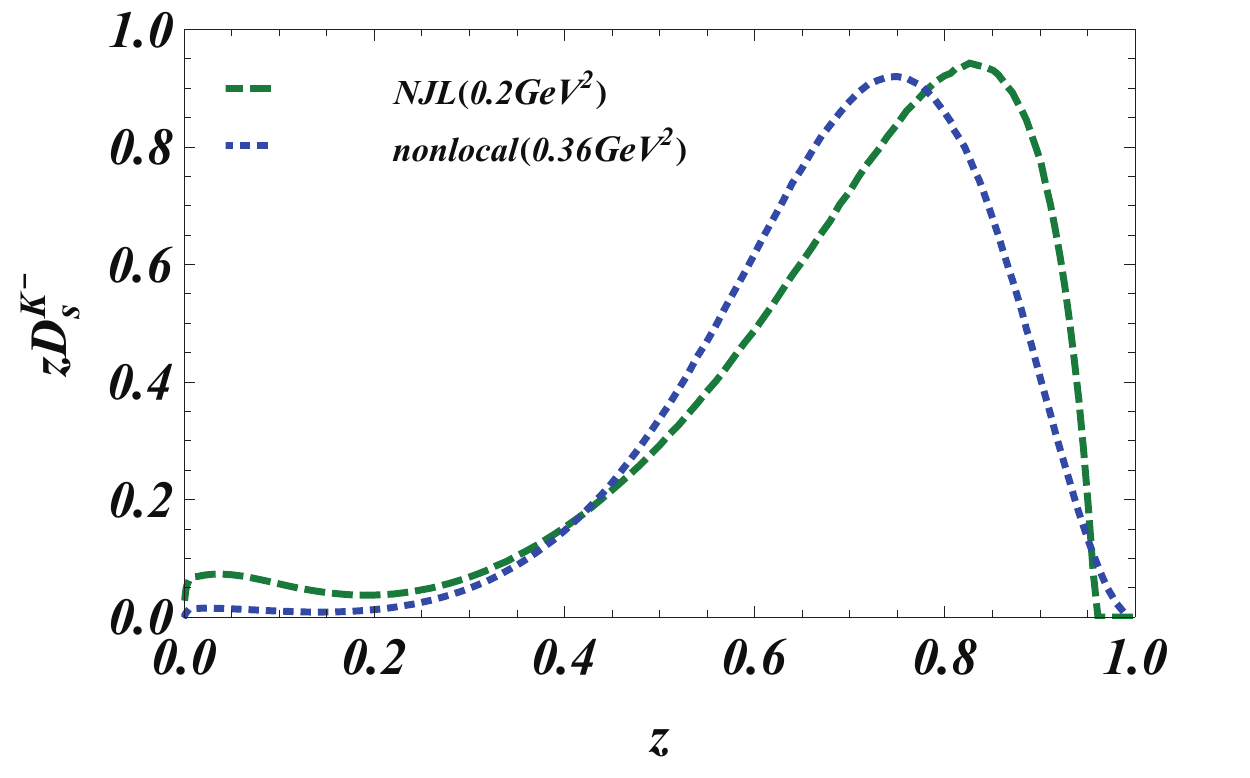}
\includegraphics[width=8.5cm]{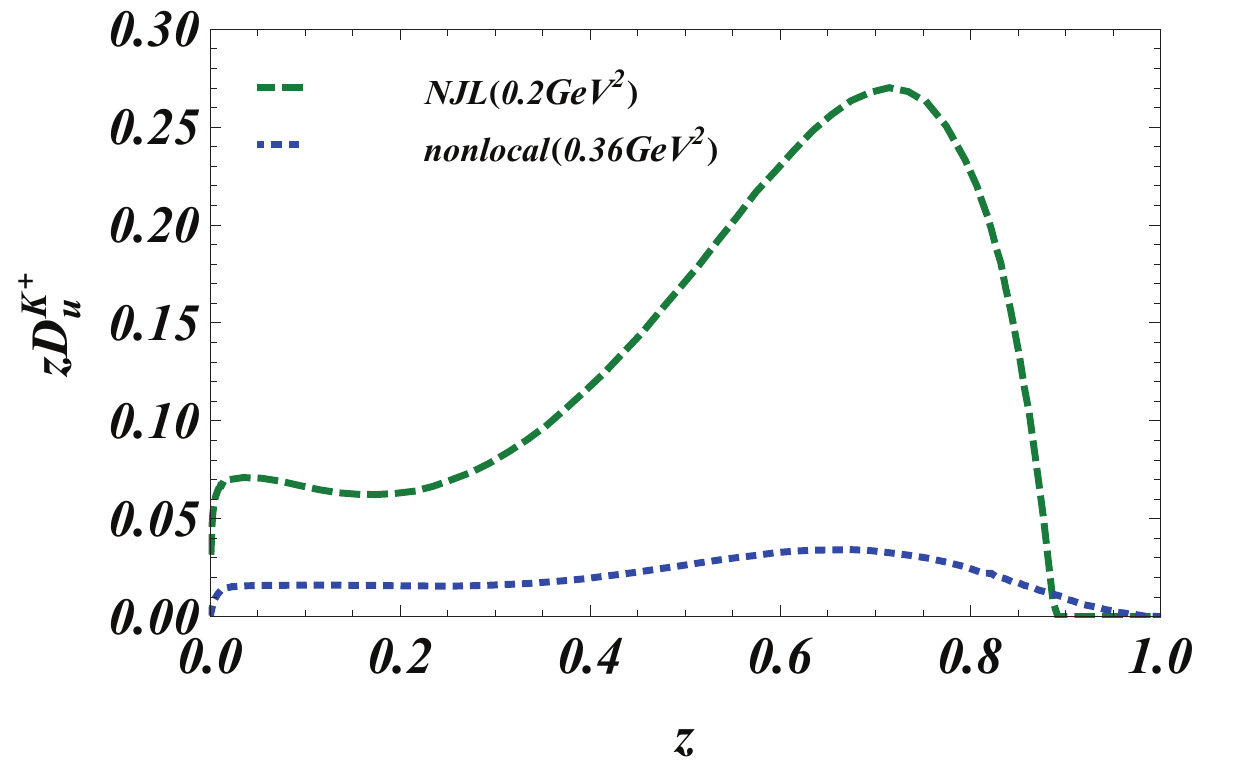}
\end{tabular}
\caption{The fragmentation functions of $zD^{\pi^{+}}_{u}(z)$ (upper panel, left), $zD^{\pi^{0}}_{u}(z)$ (upper panel, right),
$zD^{K^{-}}_{s}(z)$ (bottom panel, left) and $zD^{K^{+}}_{u}(z)$ (bottom panel, right). The dashed lines denote the results of NJL model.
The solid lines stand for the results of nonlocal chiral quark model employed in our calculations. }
\label{A1}
\end{figure}
To calculate the quark-jet contribution to the fragmentation functions within our model, we follow the approach in
Refs.~\cite{Ito:2009zc,Matevosyan:2010hh,Matevosyan:2011ey,Matevosyan:2011vj}. The elementary fragmentation functions
$\hat{d}^{h}_{q}(z)$ are re-defined as follows,
\begin{equation}
\sum_{h}\int\hat{d}^{h}_{q}(z)=\sum_{Q}\int\hat{d}^{Q}_{q}(z)dz=1,
\end{equation}
where the complementary fragmentation functions $\hat{d}^{Q}_{q}(z)$ are given by
\begin{equation}
\hat{d}^{Q}_{q}(z)=\hat{d}^{h}_{q}(1-z).\,\,\,\,\,\, h=q\bar{Q}.
\end{equation}
The fragmentation functions $D^{h}_{q}(z)$ should satisfy the following integral equation:
\begin{equation}
D^{h}_{q}(z)dz=\hat{d}^{h}_{q}(z)dz+\sum_{Q}\int^{1}_{z}dy\hat{d}^{Q}_{q}(y)D^{h}_{Q}\left(\frac{z}{y}\right)\frac{dz}{y}.
\label{multijet}
\end{equation}
Note that $D^{h}_{q}(z)dz$ in  Eq.~(\ref{multijet})
has an interpretation: $D^{h}_{q}(z)dz$ is the probability for a quark $q$ to emit a hadron
which carries the light-cone momentum fraction from $z$ to $z+dz$.
Since $\hat{d}^{Q}_{q}(y)dy$ is the probability for a quark $q$ to emit a hadron with flavor composition $q\bar{Q}$ at one step and the final quark becomes
$Q$ with the light-cone momentum fraction from $y$ to $y+dy$. Eq.~(\ref{multijet}) actually describes a fragmentation cascade process of hadron
emissions of a single quark depicted in Fig.(\ref{cascade}).

To solve the coupled integral equations Eq.~(\ref{multijet}), we apply the MC method developed in Ref.~\cite{Matevosyan:2011ey}. Namely, we simulate the
fragmentation cascade of a quark $N_\mathrm{tot}$ times and each time the fragmentation cascade stops after the quark emits
$N_\mathrm{links}$ hadrons. The fragmentation function $D^{h}_{q}(z)$ is then extracted through
the average number of type $h$ hadron with light-cone momentum fraction $z$ to $z+\Delta z$, $N^{h}_{q}(z,z+\Delta z)$,
by
\begin{equation}
D^{h}_{q}(z)\Delta z
=\frac{1}{N_\mathrm{tot}}\sum_{N_\mathrm{tot}}N^{h}_{q}(z,z+\Delta z).
\end{equation}

\begin{figure}[t]
\begin{tabular}{c}
\includegraphics[width=8.5cm]{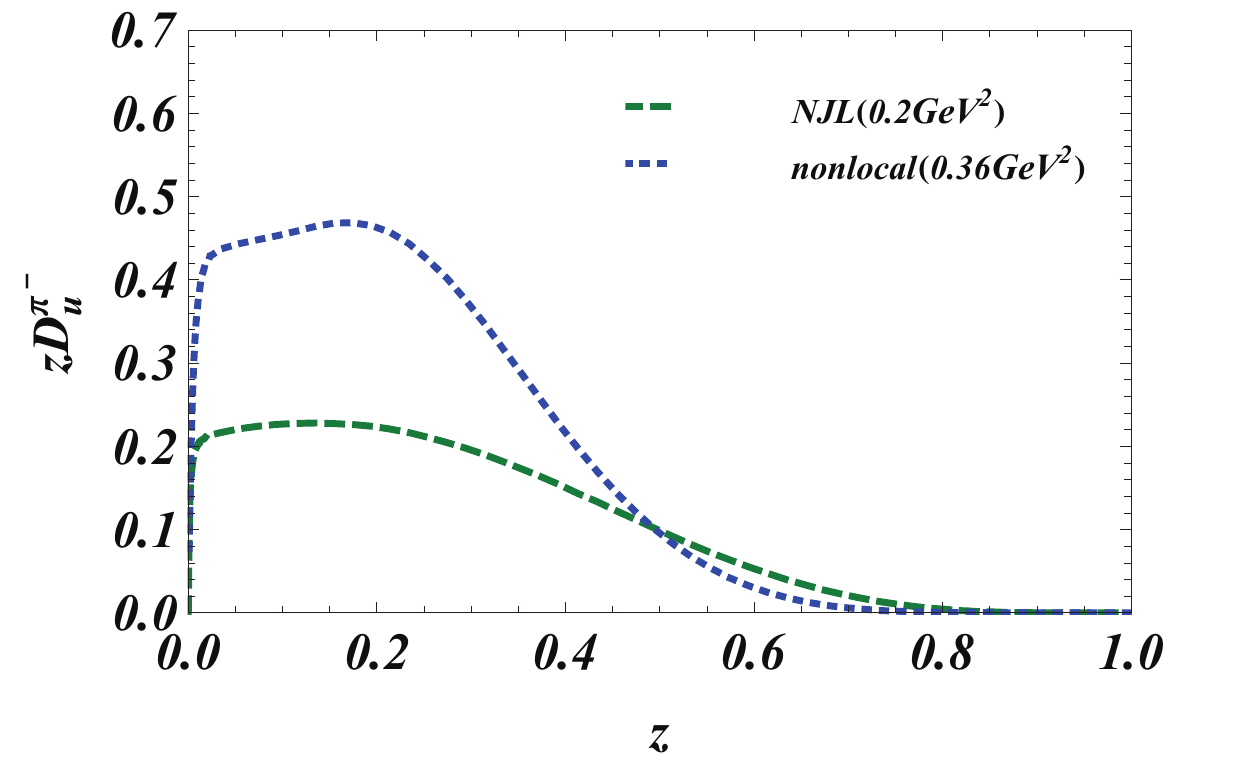}
\end{tabular}
\begin{tabular}{cc}
\includegraphics[width=8.5cm]{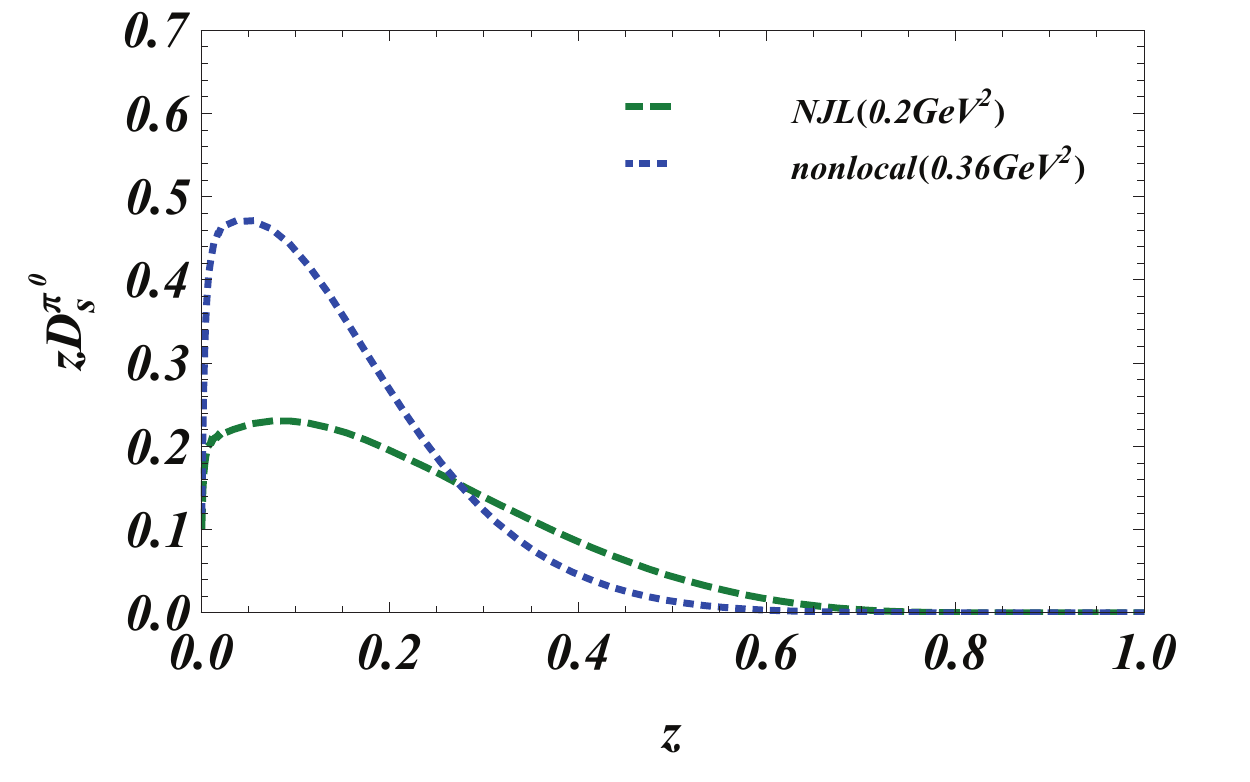}
\includegraphics[width=8.5cm]{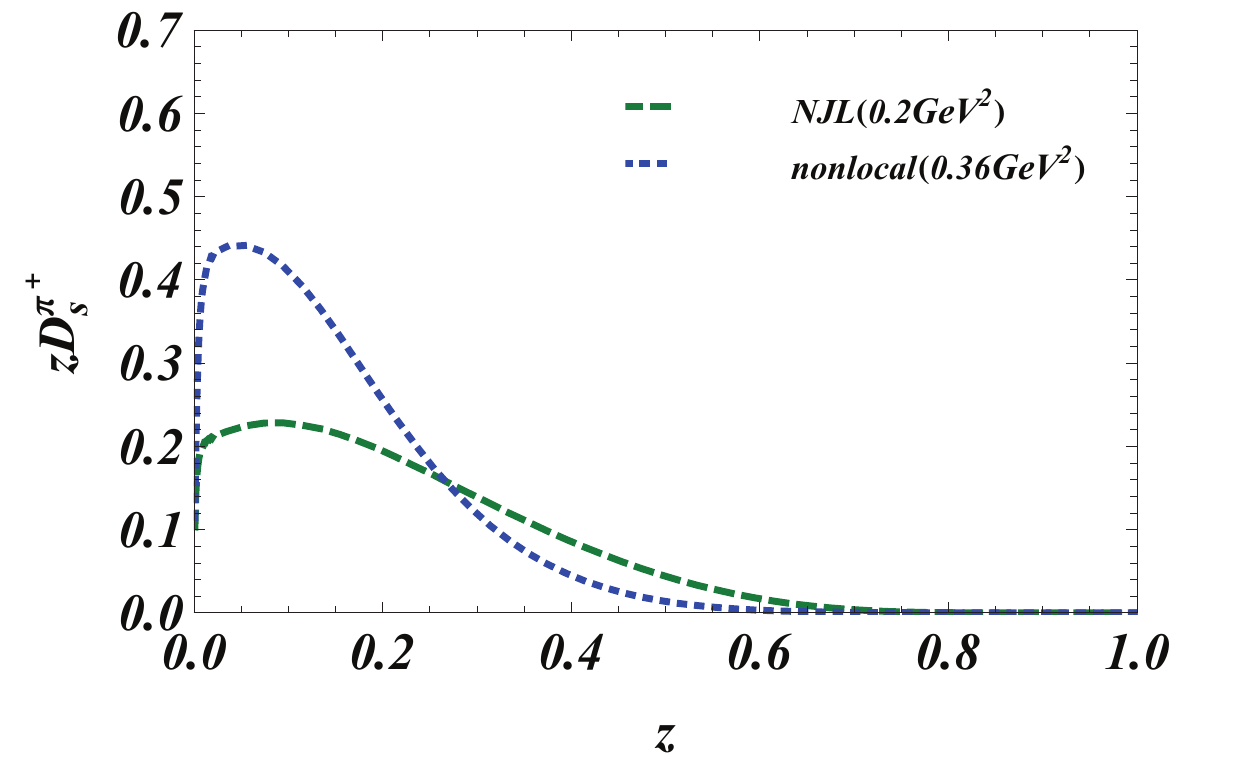}
\end{tabular}
\caption{The fragmentation functions of $zD^{\pi^{-}}_{u}(z)$ (upper panel), $zD^{\pi^{0}}_{s}(z)$(down panel, left) and $zD^{\pi^{+}}_{s}(z)$(down panel, right).
The dashed lines denote the result of NJL model. The solid lines stand
for the results of nonlocal chiral quark model employed in our calculations.}
\label{A2}
\end{figure}
When $N_\mathrm{links}$ increases, one finds that $D^{h}_{q}(z)\Delta z$ increases in the low $z$ regime.
This is due to the fact that when more steps of a fragmentation cascade are considered, more hadrons with low $z$ are emitted.
One interesting feature is that when $z\rightarrow 0$, $zd^{h}_{q}(z)\rightarrow 0$ but $zD^{h}_{q}(z)\rightarrow$ constant.
This is true for both of the NJL-jet model and our model. The value of $D^{h}_{q}(z)\Delta z$ becomes insensitive to the value of $N_\mathrm{tot}$ and
$N_\mathrm{links}$ when $N_\mathrm{tot}$ and $N_\mathrm{links}$ are large enough, implying that the result of the MC simulation converges to the solution
of Eq.~(\ref{multijet}).
When $N_\mathrm{links}$ reaches 8 the result is already convergent in the case of the NJL model ~\cite{Matevosyan:2011ey}.
However for our model the results
start to converge as $N_\mathrm{links}\ge 15$.
This can be explained by the fact that in the NJL model the peaks of the elementary fragmentation functions
occur at higher $z$ value than in NLChQM. It indicates that the probability of a quark carrying medium momentum fraction
(0.4$\le z\le $0.8) emitting a hadron is larger in our model than in the NJL model.
Hence the MC simulation of NLChQM needs the larger value of $N_\mathrm{links}$ to reach convergence.
\begin{figure}
\begin{tabular}{cc}
\includegraphics[width=8.5cm]{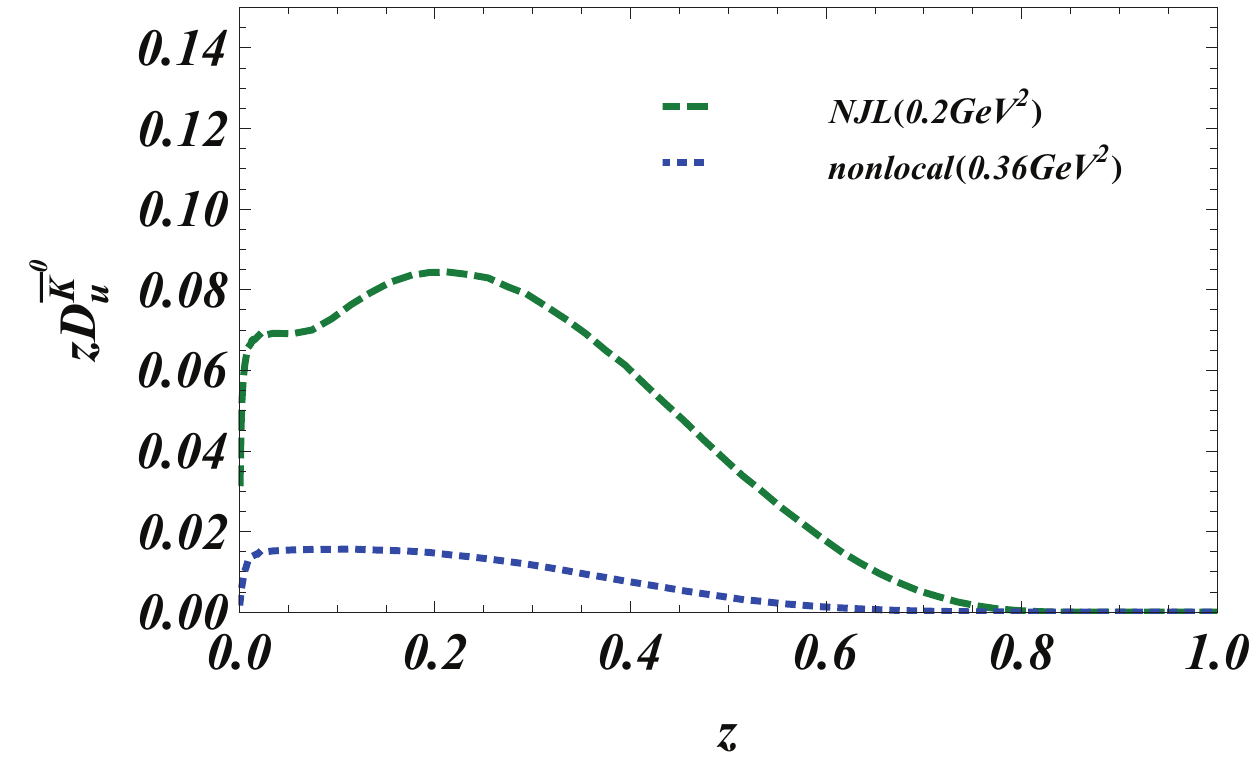}
\includegraphics[width=8.5cm]{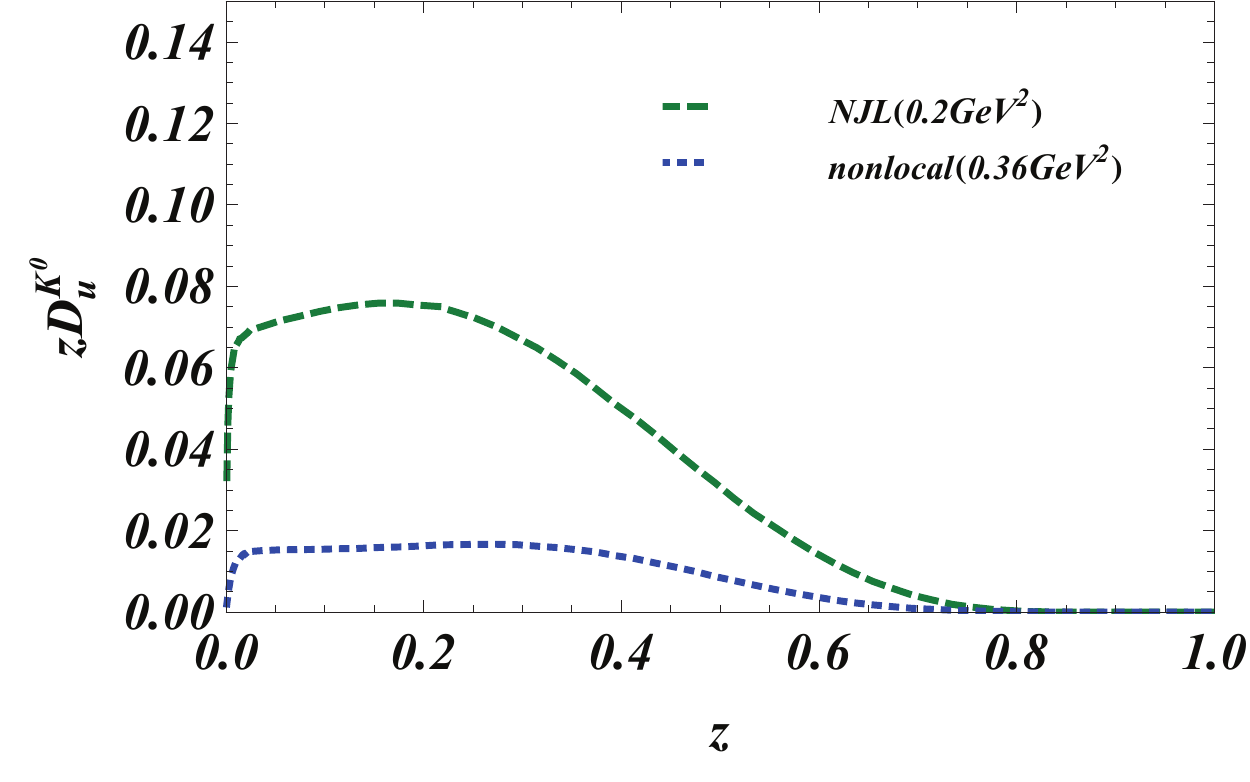}
\end{tabular}
\begin{tabular}{cc}
\includegraphics[width=8.5cm]{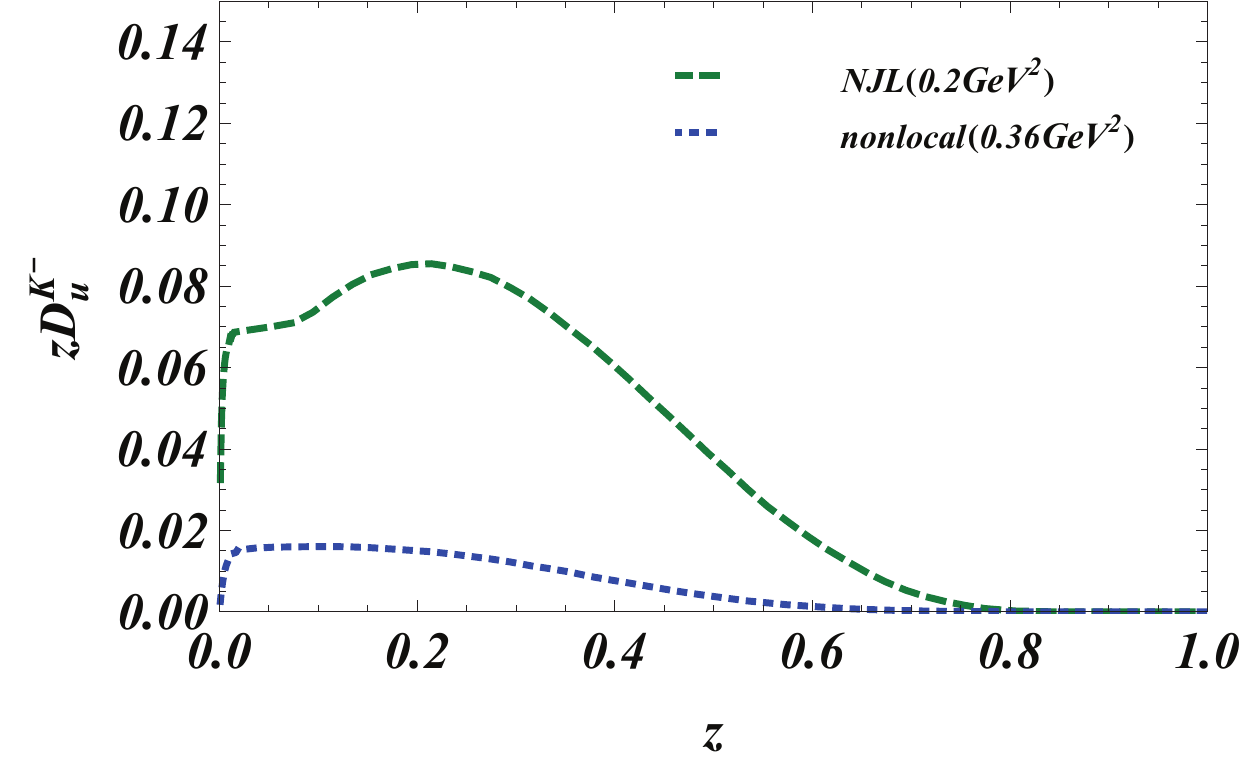}
\includegraphics[width=8.5cm]{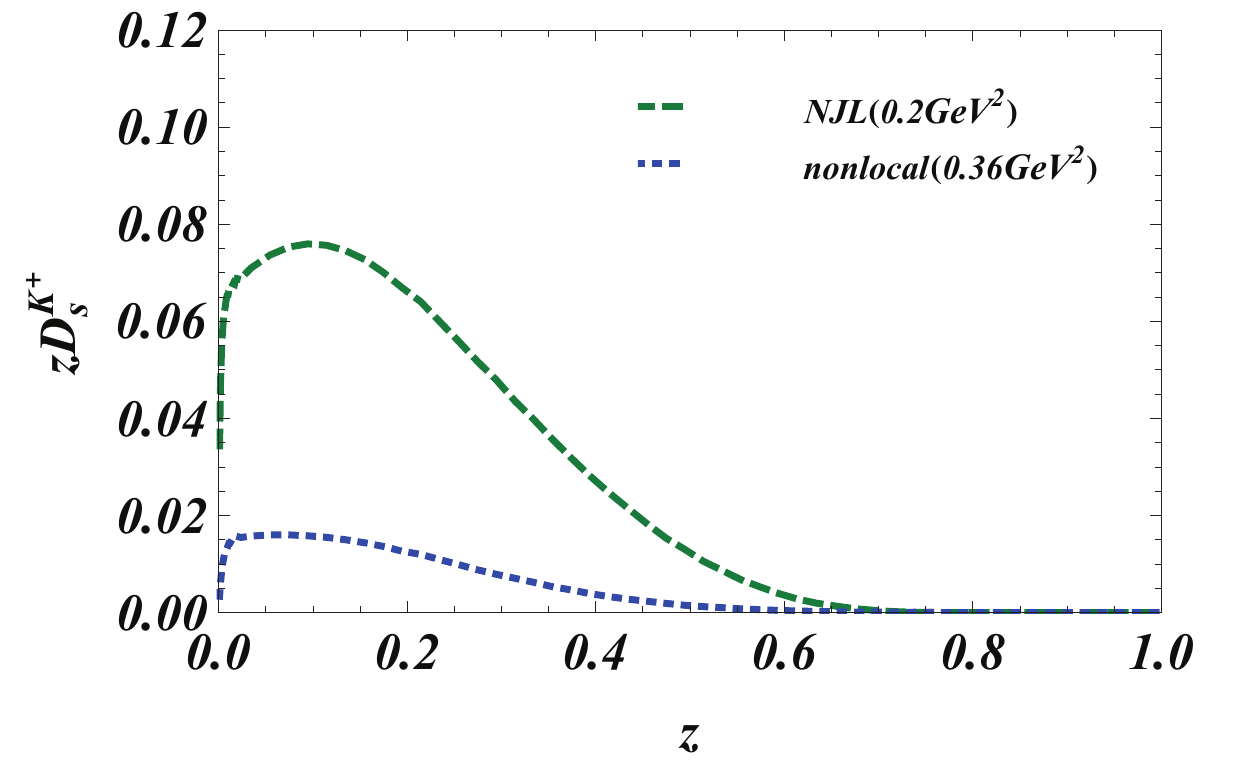}
\end{tabular}
\caption{The fragmentation functions of $zD^{K^{0}}_{u}(z)$ (upper panel, left), $zD^{\bar{K^{0}}}_{u}(z)$ (upper panel, right),
$zD^{K^{-}}_{u}(z)$ (bottom panel, left), and $zD^{K^{+}}_{s}(z)$ (bottom panel, right). The dashed lines denote the results of NJL model.
The solid lines stand for the results of nonlocal chiral quark model employed in our calculations. }
\label{A3}
\end{figure}

In this article the initial quark is only limited to be the light quark, namely, $q=u,d,s$.
In addition, the fragmented hadrons only contain pions and kaons. Naively one has 42 fragmentation functions.
According to charge conjugation and isospin symmetry, there are only 11 independent ones.
Notice among these 11 elementary fragmentation functions, only four of them are not zero.
We call these {\it direct} fragmentation functions:
\begin{eqnarray}
D^{\pi^{+}}_{u}(z)&=&D^{\pi^{-}}_{d}(z)=D^{\pi^{-}}_{\bar{u}}(z)=D^{\pi^{+}}_{\bar{d}}(z),\,\,\,
D^{\pi^{0}}_{u}(z)=D^{\pi^{0}}_{d}(z)=D^{\pi^{0}}_{\bar{u}}(z)=D^{\pi^{0}}_{\bar{d}}(z),\nonumber \\
D^{K^{+}}_{u}(z)&=&D^{K^{0}}_{d}(z)=D^{K^{-}}_{\bar{u}}(z)=D^{\overline{K^{0}}}_{\bar{d}}(z),\,\,\,
D^{K^{-}}_{s}(z)=D^{\overline{K^{0}}}_{s}(z)=D^{K^{+}}_{\bar{s}}(z)=D^{K^{0}}_{\bar{s}}(z).
\nonumber
\label{eq:D1}
\end{eqnarray}
In Fig.~\ref{A1} we present $zD^{\pi^{+}}_{u}(z)$, $zD^{\pi^{0}}_{u}(z)$, $zD^{K^{+}}_{u}(z)$ and $zD^{K^{-}}_{s}(z)$ at $\mu^2=0.36\,\mathrm{GeV}^2$.
The solid lines represent the nonlocal chiral quark model (NLChQM) results and the dotted lines stand for the corresponding
calculations determined from the NJL-jet model.

Here we briefly discuss the main features of the results of our model and the NJL-jet model.
The direct fragmentation functions ones are presented in Fig.~\ref{A1}.
In the case of $u\rightarrow \pi^{+}$, we find that the shapes of
the curves of these two models are very different. The peak of our curve occurs at $z$=0.5, but the peak of the NJL-jet curve takes place at
$z=0.8$. When $z\ge 0.6$ our result is smaller than the NJL-jet one. Between $z=0.4$ and $z=0.6$, NLChQM result increases as $z$ decreases but the
NJL-jet curves decreases. Below $z$=0.4 the two curves behave similarly but the magnitude of the NLChQM result is about twice of the NJL-jet result.
Between $z=0$ and $z=0.2$ there are plateaus for the both curves.
In our model, the possibility for the fragmented pion carrying a small momentum
fraction ($z\le 0.4$) is much larger than in the NJL-jet model.
Another direct fragmentation function of pions is $D^{\pi^{0}}_{u}(z)$.
The elementary fragmentation function $d^{\pi^{0}}_{u}(z)$ is exactly one half of $d^{\pi^{+}}_{u}(z)$,
$d^{\pi^{0}}_{u}(z)$=$\frac{1}{2}d^{\pi^{+}}_{u}(z)$.
After including the quark-jet contribution the shapes of $zD^{\pi^{0}}_{u}$ and $zD^{\pi^{+}}_{u}$ become quite
different. Our $zD^{\pi^{0}}_{u}$ is quite flat between $0\le z\le 0.4$, then it decreases as $z$ increase as $z\ge 0.4$.
On the contrary, our $zD^{\pi^{+}}_{u}$ is flat between $0\le z\le 0.2$, then it increases as $z$ increases till $z \sim 0.4$.
For $z\ge 0.4$ the curve of $zD^{\pi^{+}}_{u}$ decreases as $z$ increases.
It shows that the effect of including the quark-jet is very pronounced in our model compared to the NJL-jet model.
It is worth mentioning that the relation $zD^{\pi^{0}}_{u}\approx \frac{1}{2}D^{\pi^{+}}_{u}$ is held for $z\ge 0.6$ for both of the models.
This can be explained as follows:
since it is difficult for a quark which has already emitted hadrons in the cascade process to be fragmented into a hadron with high momentum fraction,
the quark-jet contribution is less important at high $z$ regime.
For the fragmentation process of $s\rightarrow K^{-}$, we find that
the quark-jet contribution is small for both of the models. The value of $zD^{K^{-}}_{s}$
in the regime $z\le 0.2$ is negligible. However, for the fragmentation process of $u\rightarrow K^{+}$, it is clear that our result is much
smaller than the NJL-jet result. It is of no surprise because our elementary fragmentation function $d^{K^{-}}_{u}(z)$ is very small.\\
From Table~\ref{TABLE0}, there are several elementary fragmentation functions, such as $d^{\pi^{-}}_{u}$ and $d^{\pi^{+}}_{s}$, are identically zero.
After including the quark-jet contributions, those fragmentation functions are no longer zero. We call these fragmentation functions {\it indirect}
fragmentation functions and they are generated from the process of fragmentation cascade (i.e. Fig.~\ref{cascade}).
Those indirect fragmentation functions are listed as follows:
\begin{eqnarray}
D^{\pi^{-}}_{u}(z)&=&D^{\pi^{+}}_{d}(z)=D^{\pi^{+}}_{\bar{u}}(z)=D^{\pi^{-}}_{\bar{d}}(z),
D^{K^{-}}_{u}(z)=D^{\overline{K^{0}}}_{d}(z)=D^{K^{+}}_{\bar{u}}(z)=D^{K^{0}}_{\bar{d}}(z),\nonumber \\
D^{K^{0}}_{u}(z)&=&D^{K^{+}}_{d}(z)=D^{\overline{K^{0}}}_{\bar{u}}(z)=D^{K^{-}}_{\bar{d}}(z),
D^{\overline{K^{0}}}_{u}(z)= D^{K^{-}}_{d}(z)=D^{K^{0}}_{\bar{u}}(z)=D^{K^{+}}_{\bar{d}}(z),\nonumber \\
D^{K^{+}}_{s}(z)&=&D^{K^{0}}_{s}(z)=D^{K^{-}}_{\bar{s}}(z)=D^{\overline{K^{0}}}_{\bar{s}}(z),
D^{\pi^{+}}_{s}(z)= D^{\pi^{-}}_{s}(z)=D^{\pi^{-}}_{\bar{s}}(z)=D^{K^{+}}_{\bar{s}}(z),
D^{\pi^{0}}_{s}(z)=D^{\pi^{0}}_{\bar{s}}(z).
\nonumber
\label{eq:D2}
\end{eqnarray}
For the indirect fragmentation functions of pions depicted in Fig.~\ref{A2},
we first observe that in NLChQM the shape of $zD^{\pi^{-}}_{u}(z)$ is somehow similar to $zD^{\pi^{0}}_{u}(z)$.
However, the plateau of the $zD^{\pi^{-}}_{u}(z)$ ($0\le z \le 0.2$) is only half of one for $zD^{\pi^{0}}_{u}(z)$ ($0\le z \le 0.4$).
The magnitudes of $zD^{\pi^{-}}_{u}(z)$ and $zD^{\pi^{0}}_{u}(z)$ are roughly the same
at $0\le z\le 0.2$.
As $z$ increases $zD^{\pi^{-}}_{u}(z)$ decreases much faster than $zD^{\pi^{0}}_{u}$ does.
To compare to the NJL-jet model, we find that
in the low $z$ regime the NLChQM curve is almost
twice larger than the NJL-jet result, but the two curves are very close as $z\ge 0.5$.
The results of $zD^{\pi^{0}}_{s}(z)$ and $zD^{\pi^{+}}_{s}(z)$ are almost identical.
Unlike $zD^{\pi^{-}}_{u}$, these two fragmentation functions monotonically
decrease even from very small $z$. Again the NLChQM results are larger than the NJL-jet results in the regime of $0\le z\le 0.2$.
It is because the fragmented $\pi^{+}$ meson here is emitted by the multi-step processes, such as $s\to K^{-}u,\,\, u\to \pi^{+}d$.
Since the peak of $d^{K^{-}}_{s}(z)$ is around $z$=0.8,
the secondary $u$ quark most likely carries small momentum fraction $z\le 0.2$.
The value of $D^{\pi^{+}}_{u}(z=0.2)$ of NLChQM is
larger than the corresponding one in NJL-jet model.
As a result the chance of a $s$ quark to be fragmented into a pion is larger in NLChQM.\\
In Fig.~\ref{A3} we present four indirect fragmentation functions of the kaons,
$zD^{K^{0}}_{u}(z)$, $zD^{\bar{K^{0}}}_{u}(z)$, $zD^{K^{-}}_{u}(z)$, and $zD^{K^{+}}_{s}(z)$. The common feature of NLChQM results is that they are all almost
one order magnitude smaller than the pion ones. In contrast to the pion ones, the NJL-jet results are much larger than ours.
It is because these indirect fragmentation functions are related to the kaon elementary
fragmentation functions which are tiny.
Consequently those associated indirect kaon fragmentation functions are also suppressed.
For example, the process $s\rightarrow K^{0}$ is the combination of the two processes such as $u\rightarrow \pi^{+}d,\,d\rightarrow K^{0}s$.
Since $d^{K^{0}}_{d}(z)$=$d^{K^{+}}_{u}(z)$ is very small, naturally $D^{K^{0}}_{s}$ is also very small.\\
\begin{figure}[t]
\begin{tabular}{cc}
\includegraphics[width=8.5cm]{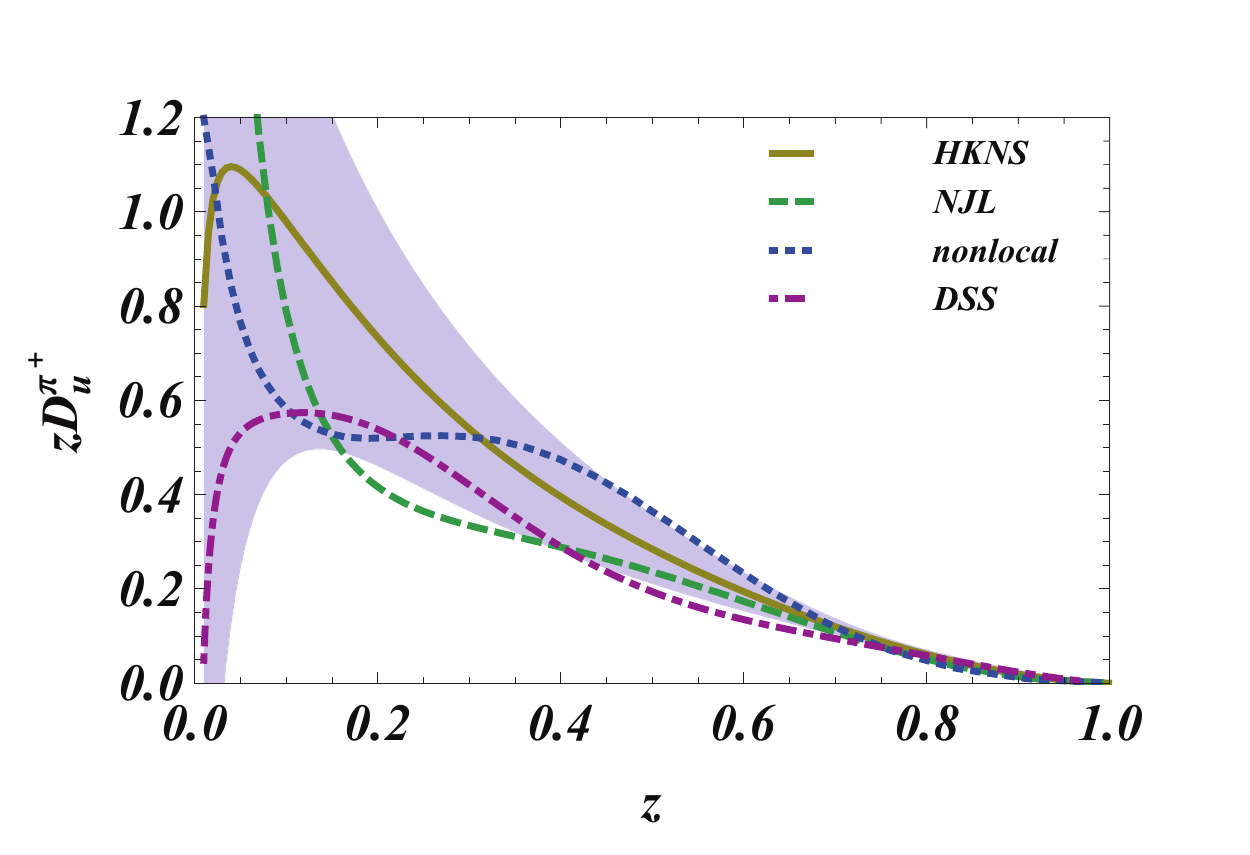}
\includegraphics[width=8.5cm]{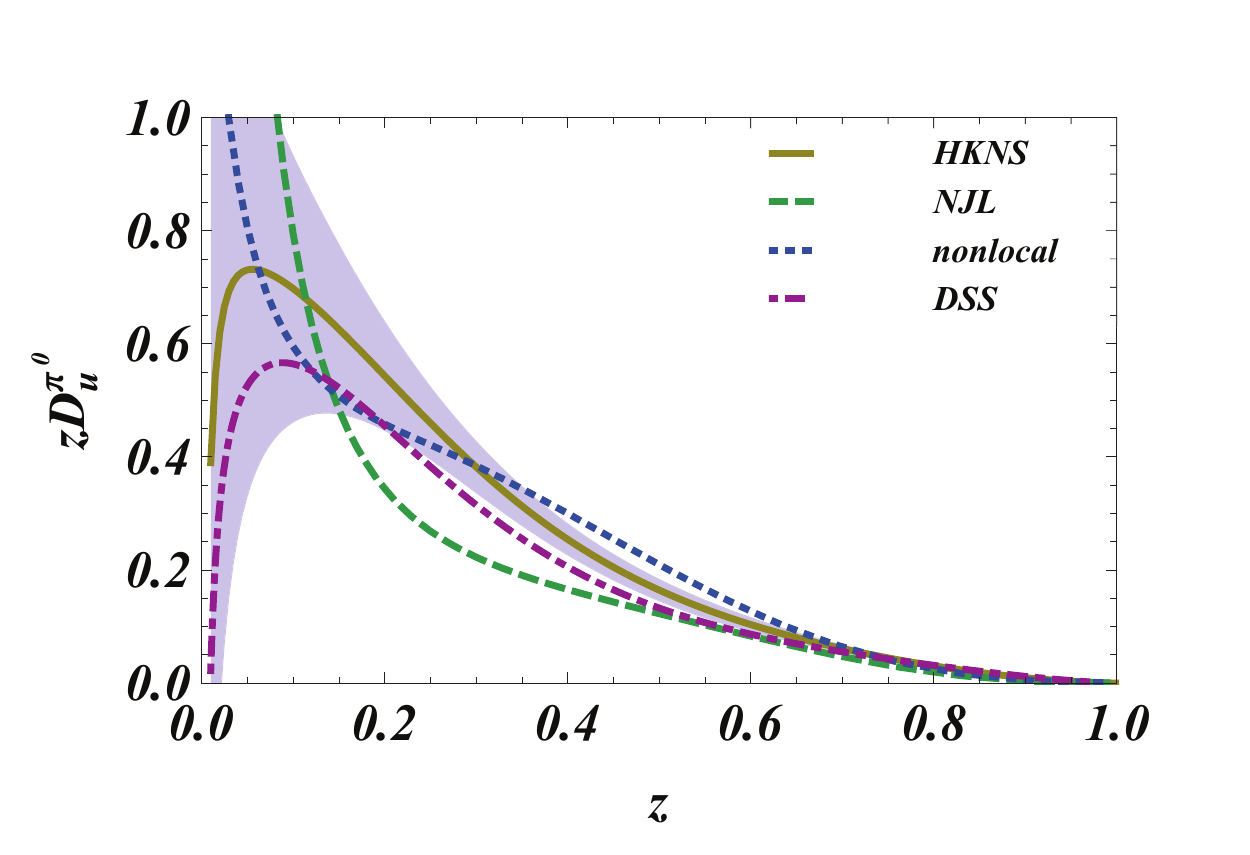}
\end{tabular}
\begin{tabular}{cc}
\includegraphics[width=8.5cm]{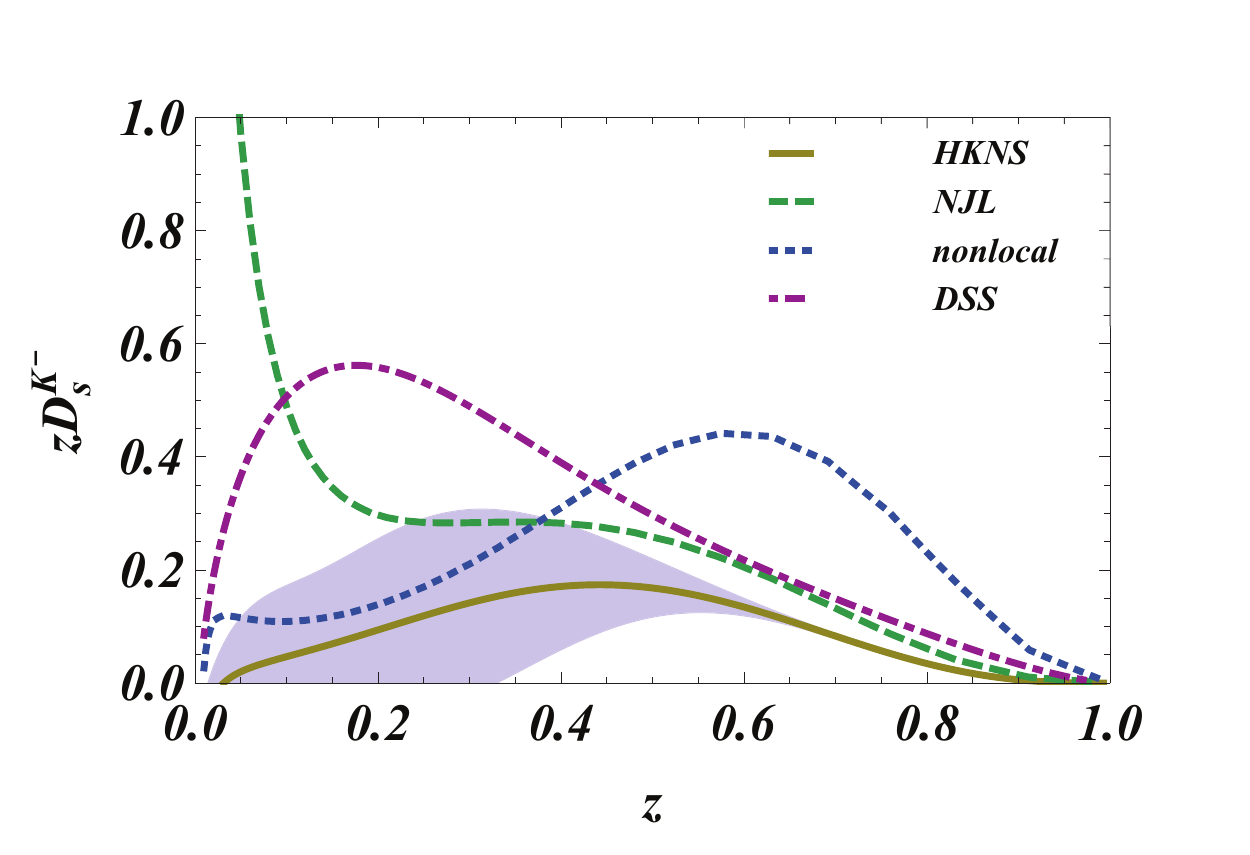}
\includegraphics[width=8.5cm]{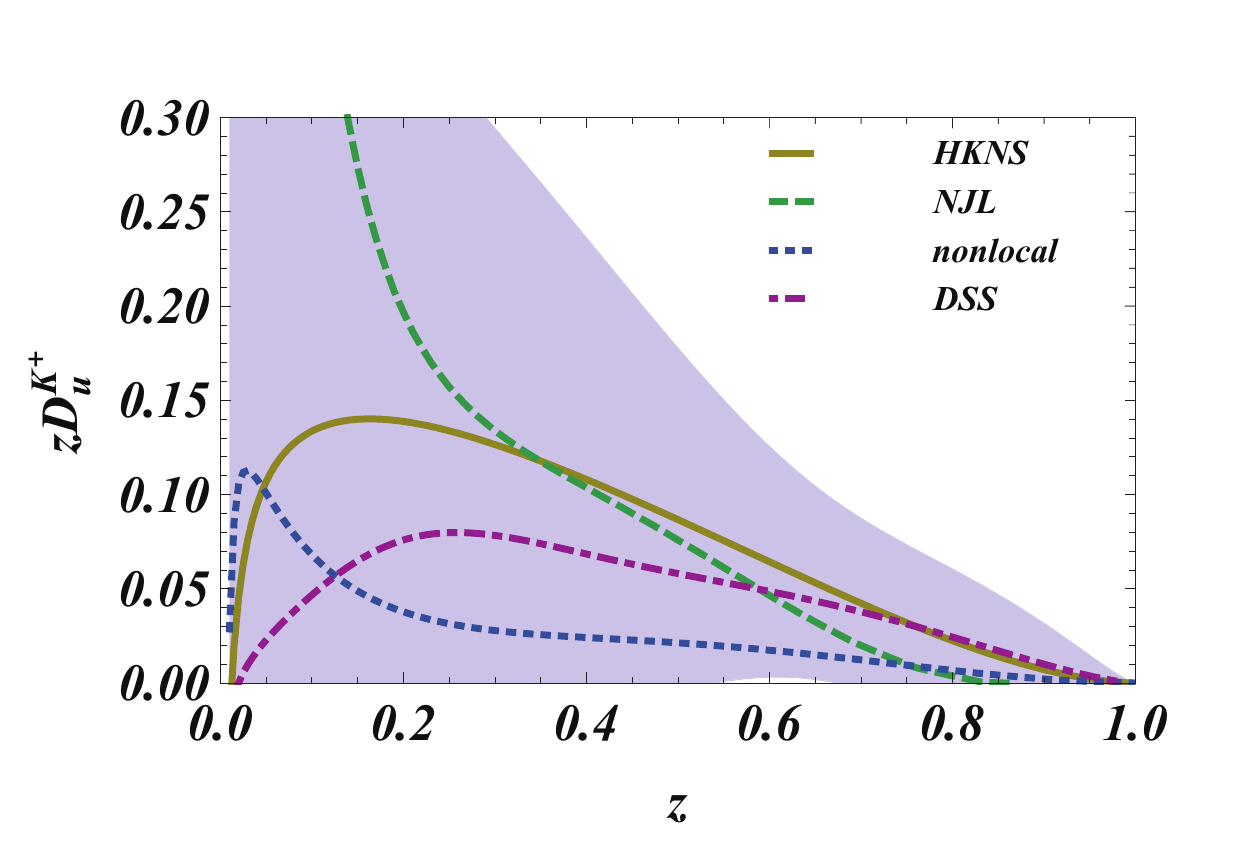}
\end{tabular}
\caption{The fragmentation functions of $zD^{\pi^{0}}_{u}(z)$ (upper panel, right) and $zD^{\pi^{+}}_{u}(z)$ (upper panel, left),
$zD^{K^{+}}_{u}(z)$ (bottom panel, right) and $zD^{K^{-}}_{s}(z)$ (bottom panel, left). While the dashed lines denote the results of NJL model,
the solid lines stand for the results of nonlocal chiral quark model. The dotted lines are the HKNS curves and the dot-dashed lines are DSS curves.
HKNS and DSS are two empirical parameterizations of the fragmentation functions. The uncertainty bands are according to HKNS parameterizations.}
\label{B1}
\end{figure}
In summary, we find that our results are substantially different from the NJL-jet model results after including the quark-jet contribution.
For pion ones the NLChQM results are higher than the NJL-jet ones in the medium and low $z$ regime.
For the kaon ones the situation is rather different.
For $s\rightarrow K^{-}$ we arrive at a result similar to that from the NJL-jet model, but our investigation implies that the process of
$u\to K^{+}$ is highly suppressed. For the other channels NLChQM results are always smaller than the NJL-jet model.
\section{Numerical Results of the fragmentation functions at $Q^2=4\,\mathrm{GeV}^2$}
In this section we will present our results at $Q^2=4\,\mathrm{GeV}^2$ and compare them with the empirical parametrizations and
the NJL-jet model results. We employ QCDNUM17~\cite{Conway:1989fs,Botje:2010ay,DGLAP} to evolute our results from $Q^2=0.36\,\mathrm{GeV}^2$ to
$Q^2=4\,\mathrm{GeV}^2$. Since $D^{\pi^{+}}_{u}(z)$ is the most pronounced process, therefore, the initial momentum for evolution is
determined by a reasonable agreement between our evolution result of $D^{\pi^{+}}_{u}(z)$ with two empirical parameterizations, namely the
HKNS parametrization~\cite{Hirai:2007cx} and the DSS parametrization ~\cite{deFlorian:2007aj}.
These two empirical parameterizations are used for comparison of other fragmentation functions as well.
\begin{figure}[t]
\begin{tabular}{c}
\includegraphics[width=8.5cm]{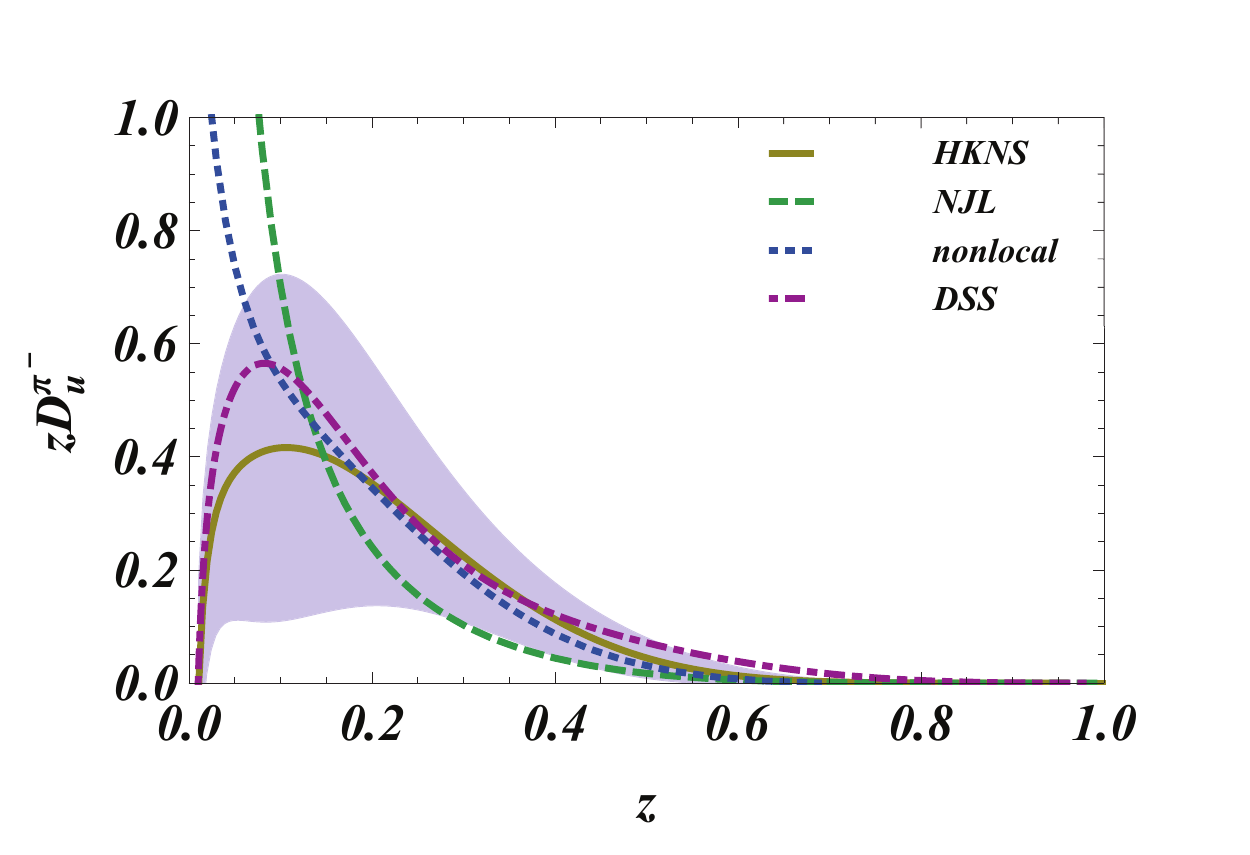}
\end{tabular}
\begin{tabular}{cc}
\includegraphics[width=8.5cm]{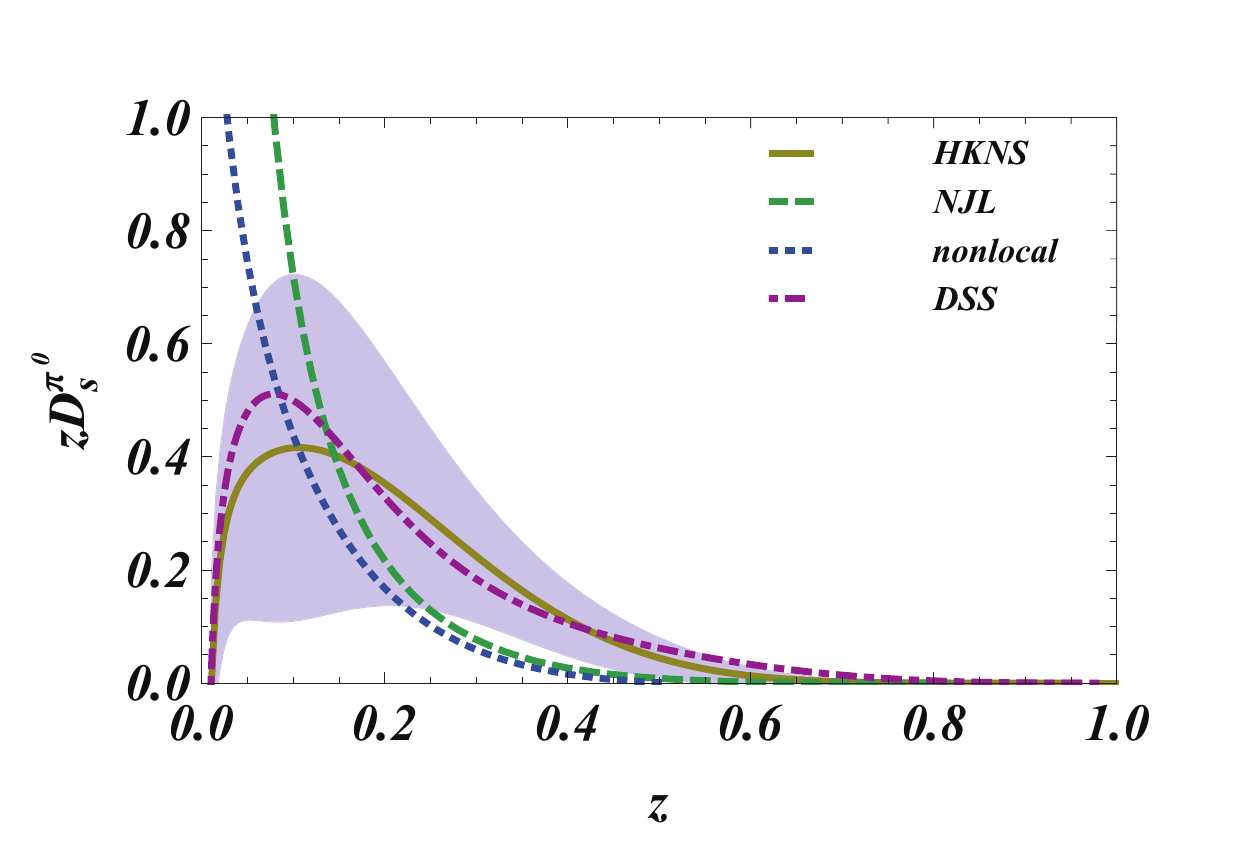}
\includegraphics[width=8.5cm]{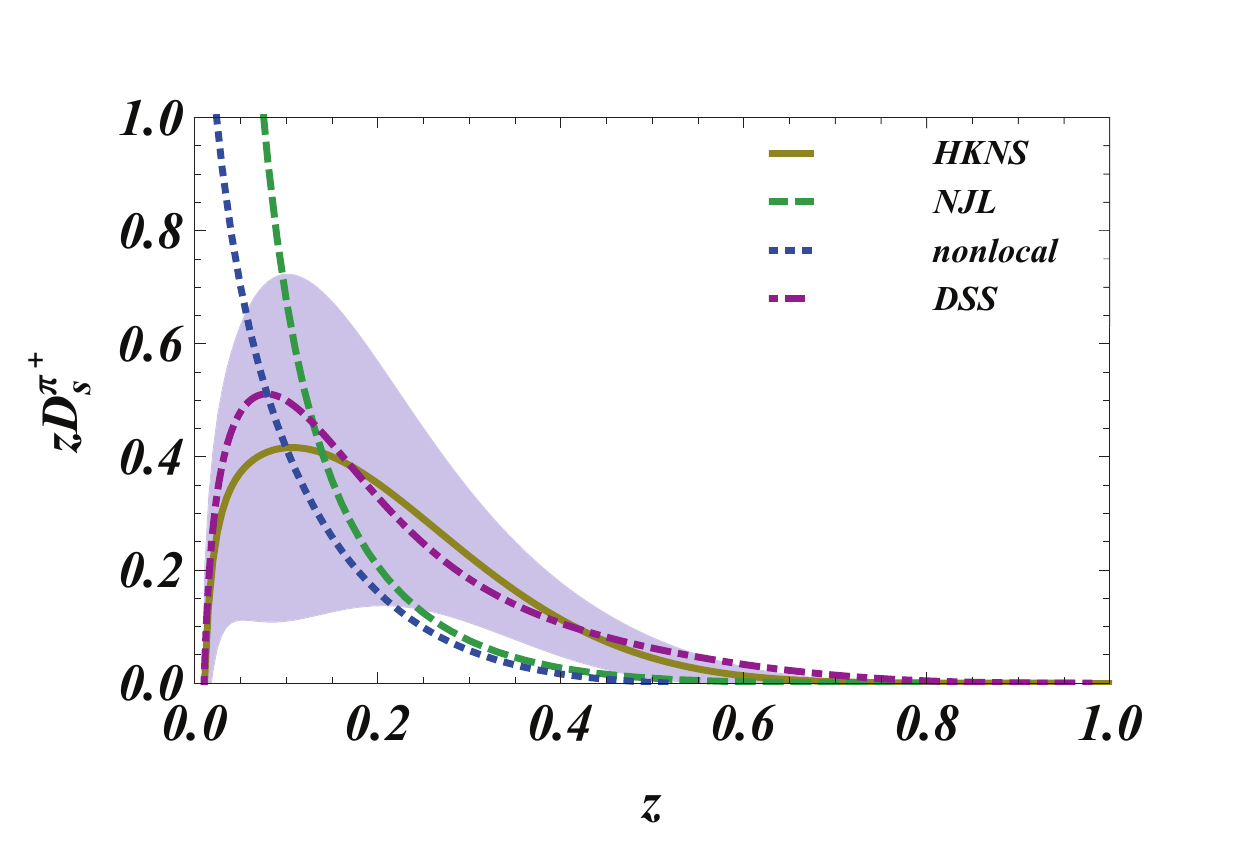}
\end{tabular}
\caption{The fragmentation functions of $zD^{\pi^{-}}_{u}(z)$ (upper panel) and $zD^{\pi^{0}}_{s}(z)$(down panel, left) and $zD^{\pi^{+}}_{s}(z)$(down panel right).
The dashed lines denote the result of NJL model. The solid lines represent the results of nonlocal chiral quark model. The dotted lines are the HKNS
curves and the dot-dashed lines are the DSS curves. HKNS and DSS are two empirical parameterizations of the fragmentation functions. The uncertainty bands
are according to the HKNS parameterizations.}
\label{B2}
\end{figure}
The results of direct fragmentation functions evolved to $Q^2=4\,\mathrm{GeV}^2$ are given in Fig.~\ref{B1}.
The dashed lines denote the results of NJL
model. In addition, the solid lines stand for the results of nonlocal chiral quark model employed in our calculations. Finally, the dotted and
dot-dashed lines are the HKNS curve and DSS curve, respectively.
The uncertainty bands are provided by the HKNS parameterization. \\
The NLChQM result of $zD^{\pi^{+}}_{u}(z)$ is within the uncertainty band of HKNS result.
In the high $z$ regime our result is consistent with the two parameterizations and the NJL-jet model. In the medium $z$ region ($0.3\le z\le 0.7$)
the NLChQM result is slight higher than HKNS and DSS. Between $z=0.1$ and $z=0.4$, it appears to be a plateau then turns up at $z=0.1$.
On the contrary, the NL-jet result is slight below the HKNS and DSS parameterizations between $z=0.2$ and $z=0.4$ and turns up at $z=0.2$.
In the case of $zD^{\pi^{0}}_{u}$, NLChQM result is slightly higher than the two parameterizations between $z$=0.3 and $z$=0.7.
On the contrary, the NJL-jet result is clearly below the uncertainty between $z$=0.2 to $z$=0.4.
For the both cases of $zD^{\pi^{+}}_{u}$ and $zD^{\pi^{0}}_{u}$, the results of NLChQM and NJL-jet agree with HNKS and DSS quite well
in the high $z$ regime $z\ge 0.7$.
We now turn our attention to the case of $zD^{K^{+}}_{u}$.
We find that both of NLChQM and NJL-jet curves are within the uncertainty band.
However, the
shapes of the two curves are completely different. The NLChQM curve increases from $z$=0 to $z$=0.05 then decreases rapidly from $0.05$ to $0.2$,
after which it decreases slowly. The NJL-jet curve simply decreases as $z$ increases. Unfortunately, neither of them catches the feature of
the empirical curves. Model results are too small compared with the empirical ones in the region of $z\ge$ 0.7. Between $z$=0.4 and 0.7,
the NJL-jet model result is comparable with two parameterizations. On the contrary,
the NLChQM result is always too small. In the low $z$ region,
NLChQM curve becomes comparable with the empirical ones but the NJL-jet model becomes far too large.
The fragmentation function $zD^{K^{-}}_{s}(z)$ is probably the most problematic one for the model calculations.
On the one hand, the NLChQM curve is too large compared with
the empirical curves in the high $z$ region, but it turns out to lie between the two empirical curves in the region of $z\le0.4$.
On the other hand, the NJL-jet curve agrees with DSS curve in the high $z$ region quite well but becomes far too large as $z\le 0.1$. Both the
model curves fall outside the uncertainty band of HNKS parameterization from medium to high $z$ regime.

\begin{figure}[t]
\begin{tabular}{cc}
\includegraphics[width=8.5cm]{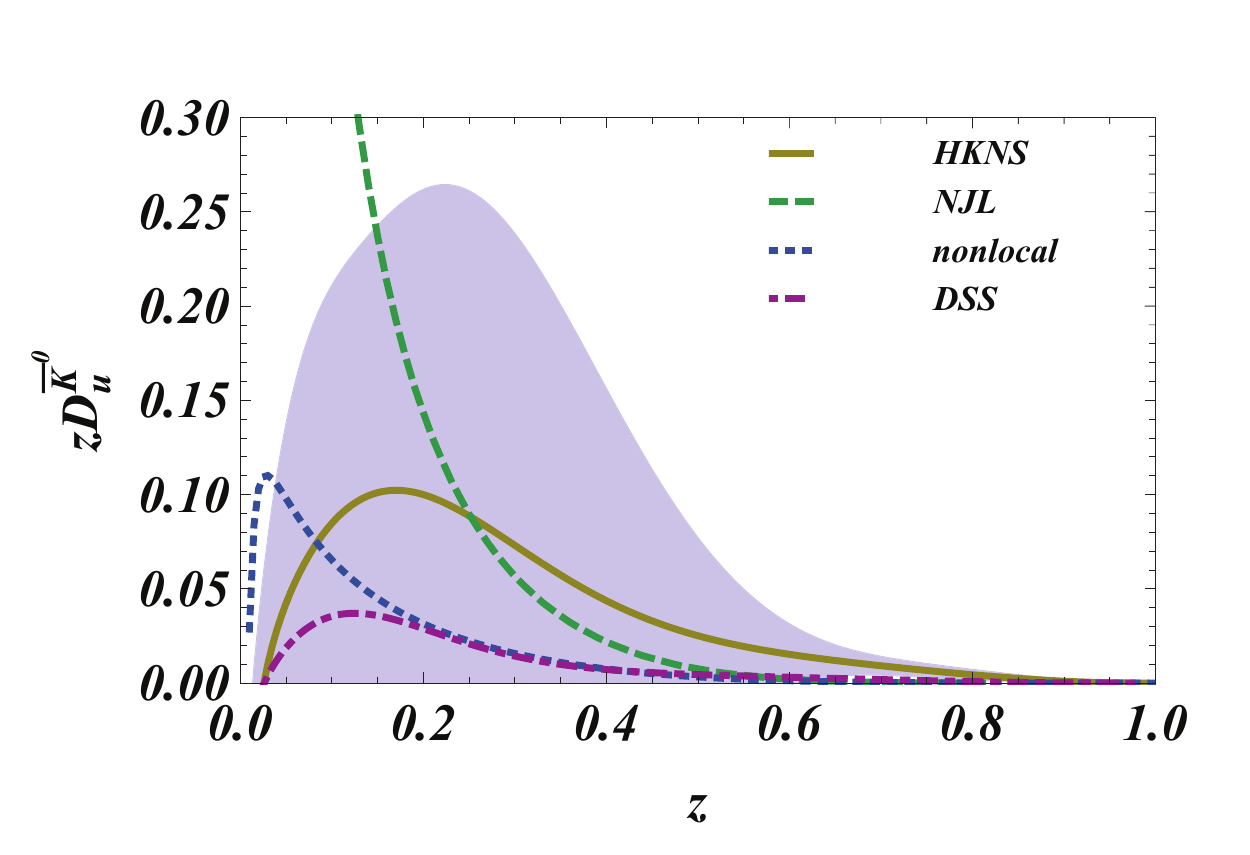}
\includegraphics[width=8.5cm]{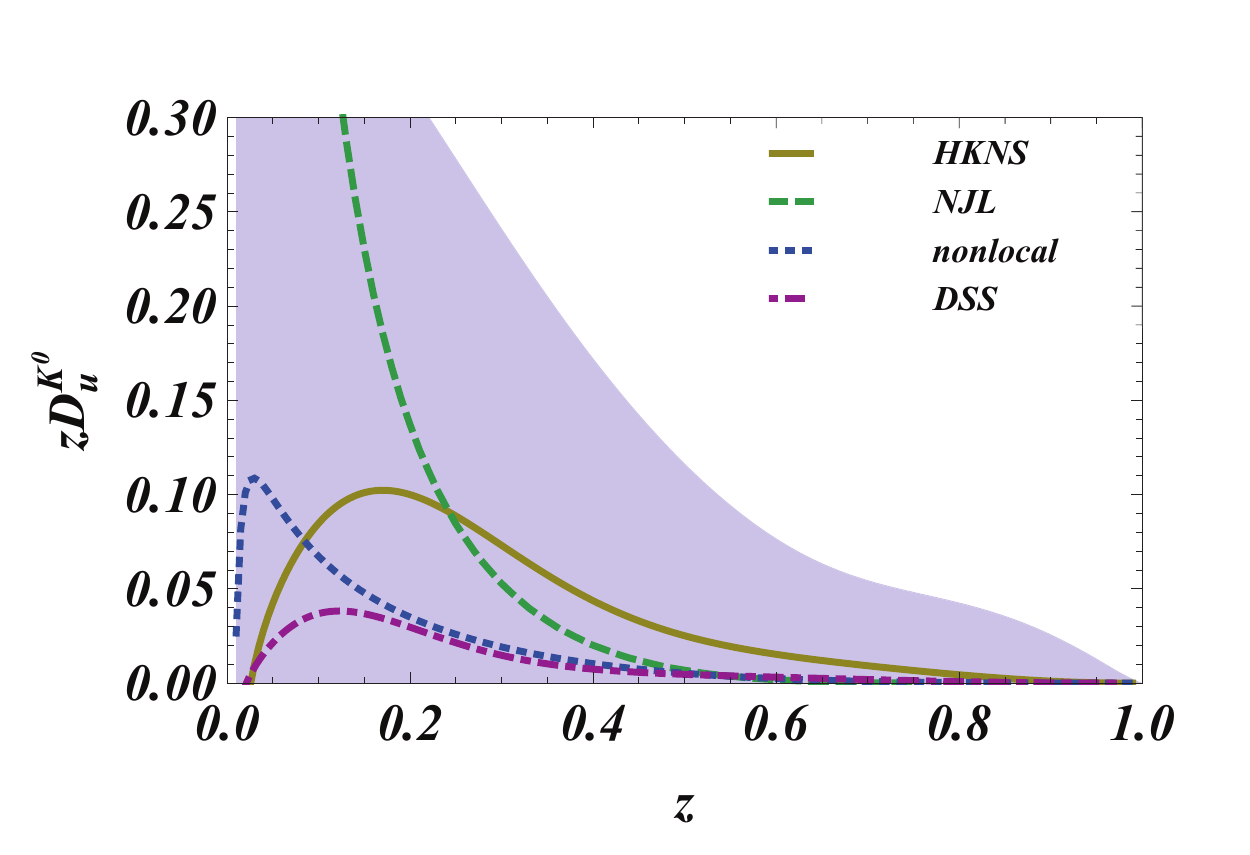}
\end{tabular}
\begin{tabular}{cc}
\includegraphics[width=8.5cm]{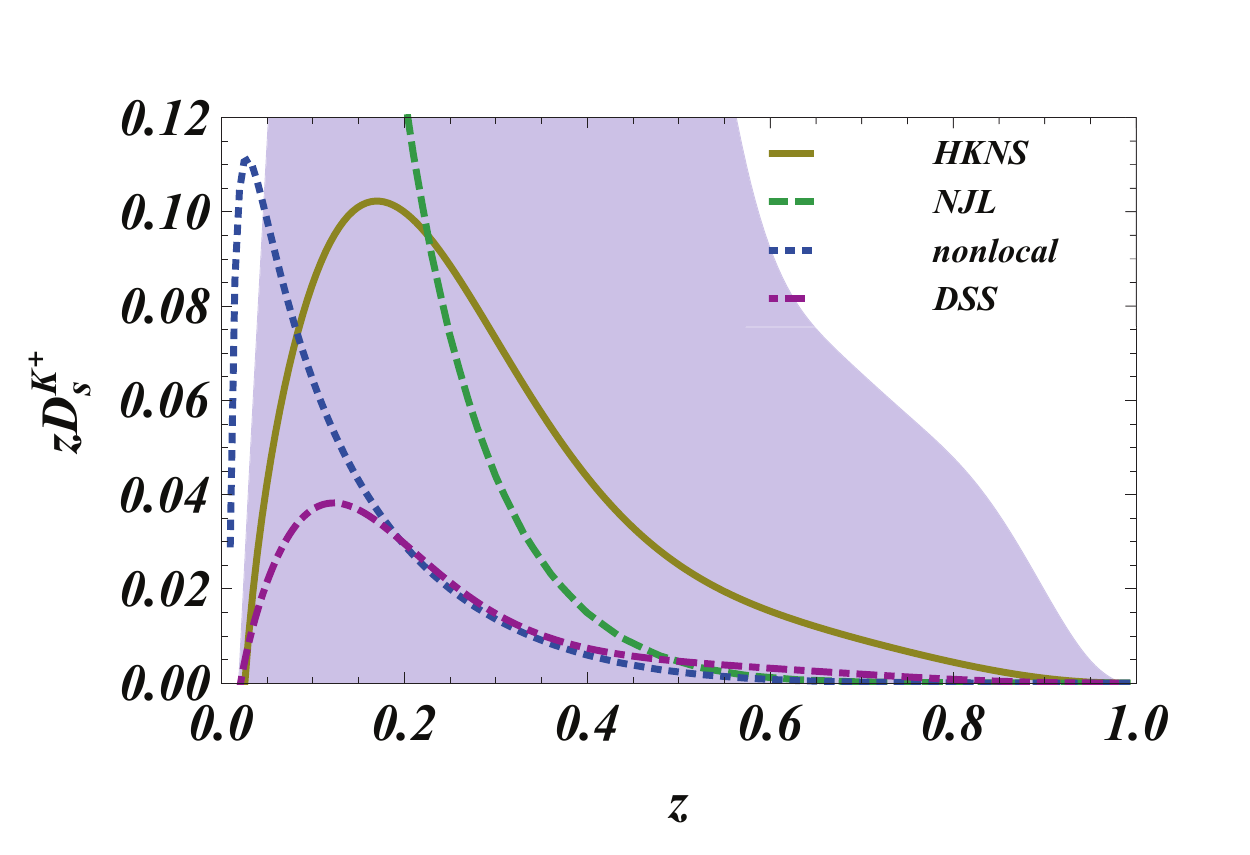}
\includegraphics[width=8.5cm]{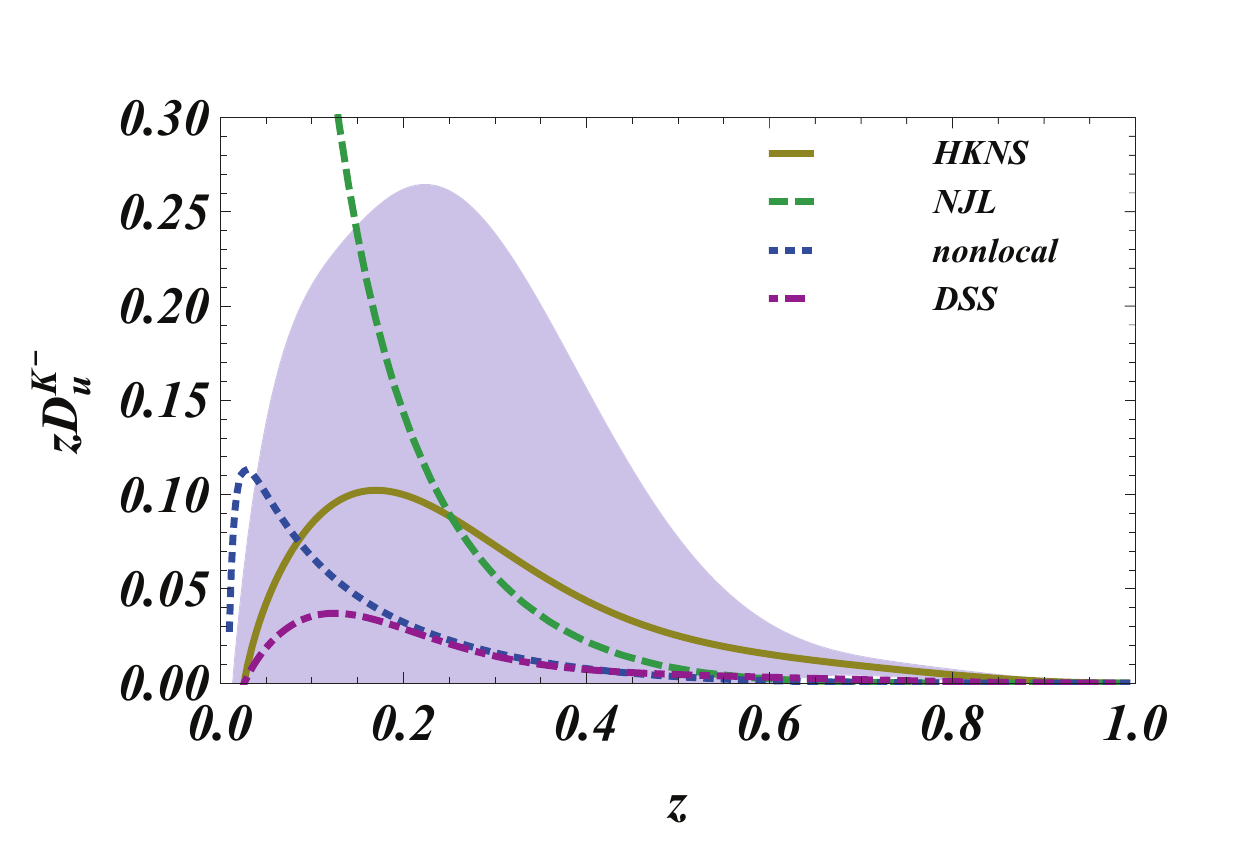}
\end{tabular}
\caption{The fragmentation functions of $zD^{K^{0}}_{u}$ (upper panel, right) and $zD^{\bar{K^{0}}}_{u}$ (upper panel, left), $zD^{K^{-}}_{u}$ (bottom panel, right),
and $zD^{K^{+}}_{s}$ (bottom panel, left). The dashed lines denote the results of NJL model. The solid lines represent the results of nonlocal chiral quark
model. The dotted and dot-dashed lines are the HKNS curves and DSS curves, respectively. HKNS and DSS are two empirical parameterizations of the
fragmentation functions. The uncertainty band is according to HKNS parameterizations. }
\label{B3}
\end{figure}
Now, let us discuss the {\it indirect} fragmentation functions. The ones of pions are depicted in Fig.~\ref{B2}.
The most successful channel for NLChQM model is $zD^{\pi^{-}}_{u}$. Between $z=1$ and $z=0.2$, the NLChQM curve agrees with both of HKNs and DSS
very well. Between $z=0.1$ and $z=0.2$ the NLChQM curve still coincides with DSS and is a little bit higher than HKNS curve. This excellent agreement
disppears when $z\le 0.1$. Compared with the NLChQM curve, the NJL-jet result locates at the lower edge of the uncertainty band
in the high $z$ region
and turns upward at $z=0.3$.
For the fragmentation process of $s\to \pi$, both of NLChQM and NJL-jet models give a very similar result. However they both underestimate the
fragmentation functions in the medium and high $z$ region. The low $z$ behavior of both results also fail to catch the feature of the empirical curves.
It remains a challenge for further study.

The indirect ones for kaons are depicted in Fig.~\ref{B3}.
The NLChQM results for $zD^{K^{0}}_{u}$, $zD^{\bar{K^{0}}}_{u}$ and $zD^{K^{-}}_{u}$ are almost identical to $zD^{K^{+}}_{u}$. However, unlike the case of
$zD^{K^{+}}_{u}$, NLChQM results of $zD^{K^{0}}_{u}$, $zD^{\bar{K^{0}}}_{u}$, and $zD^{K^{-}}_{u}$ agree with DSS curves excellently between $z=0.2$ and $z=1$.
At low $z$ regime NLChQM curves overshoot a little bit but still in a reasonable good agreement to the empirical curves. Notice that our results are
all within the uncertainty band except in the extremely low $z$ region.
The last channel we discuss is $zD^{K^{+}}_{s}$. Remarkably, the result of $zD^{K^{+}}_{s}$ is almost identical to the one of $u\to K$.
Again our model agrees with DSS curve in the high and medium $z$ region and is above the empirical curves in the low $z$ regime.

\section{Summary and outlook}
In this article, we have investigated the quark-jet contribution to the fragmentation functions of the pions and the kaons
using NLChQM and evolute them to $Q^2$=4 GeV$^2$.
The current-quark masses, $(m_u,m_d,m_s)=(5,5,150)$ MeV has been used in our calculations.
We summarize the important observations in the present work as follows:
\begin{itemize}
\item  For the direct pion fragmentation functions, NLChQM results agree with the empirical data quite well except at extremely low $z$ regime.
\item  In the case of direct kaon fragmentation functions, our result of $zD^{K^{+}}_{u}$ is underestimated in the high and medium $z$ regime.
On the other hand, our result of $zD^{K^{-}}_{s}$  is overestimated in the high $z$ regime. It is out of the uncertainty band of HKNS parametrization, either. Nevertheless, our $zD^{K^{-}}_{s}$ is still between HKNS and DSS parameterizations from medium to low $z$ regime.
\item  The most successful channel for NLChQM is $u\rightarrow \pi^{-}$. The agreement between the results of NLChQM and the empirical curves is
excellent except in the extremely low $z$ regime.
\item  For other indirect pion fragmentation functions, such as $D^{\pi^{0}}_{s}$ and $D^{\pi^{+}}_{s}$, our results are mostly within the uncertainty band
except in the low $z\le0.1$ region. Besides our results for those channels are very similar to the NJL-jet ones.
\item  For the indirect kaon fragmentation functions, NLChQM results agree with one of the empirical curve, DSS parametrization, within the regime of
$0.2\le z\le 1$. Furthermore these indirect fragmentation functions lie mostly inside the uncertainty bands.
\end{itemize}

In summary, we have shown that NLChQM provides an excellent framework to calculate the unpolarized fragmentation functions.
The results agree with the
empirical parametrizations quite well in most of the channels. There are several directions to improve and extend
our current calculations. For example,
we have not taken into account the axial-current conservation in the present framework, which may become problematic for the nonlocal quark-PS meson
interactions~\cite{Nam:2006sx}. In Ref~\cite{Nam:2012vm} this contribution has been taken into account to modify the
quark distribution functions. However, it is not straightforward to include this effect into the calculations of the elementary fragmentation functions.
This work is in progress. Another issue is to include $\eta$ and $\eta'$ in the fragmented mesons. Furthermore one should also include the vector mesons
and baryons in the fragmented hadrons. We expect to continue to study more complicated fragmentation functions, such as unpolarized dihadron fragmentation
functions, Collins fragmentation functions and polarized dihadron fragmentation functions and apply our result to extract the transverse parton
distributions of the proton.

\section*{Acknowledgments}
S.i.N. is very grateful to the hospitality during his visiting National Taiwan University (NTU) with the financial support from NCTS (North) of Taiwan,
where the present work was partially performed. C.W.K. are supported by the grant NSC 99-2112-M-033-004-MY3 from National Science Council
(NSC) of Taiwan. F.J.J. and D.J.Y are partially supported by NSC of Taiwan (grant No. NSC 99-2112-M003-015-MY3).
Support of NCTS (North) of Taiwan is also acknowledged gratefully.
\section*{Appendix}
\begin{table}[h]
\begin{tabular}{c|ccccccc}
$\mathcal{C}^{h}_q$&$\pi^0$&$\pi^+$&$\pi^-$&$K^0$&$\bar{K^{0}}$&$K^+$&$K^-$\\
\hline
$u$&$1/2$&$1$&$0$&$0$&$0$&$1$&$0$\\
$d$&$1/2$&$0$&$1$&$1$&$0$&$0$&$0$\\
$s$&$0$&$0$&$0$&$0$&$1$&$0$&$1$\\
$\bar{u}$&$1/2$&$0$&$1$&$0$&$0$&$0$&$1$\\
$\bar{d}$&$1/2$&$1$&$0$&$0$&$1$&$0$&$0$\\
$\bar{s}$&$0$&$0$&$0$&$1$&$0$&$1$&$0$\\
\end{tabular}
\caption{Flavor factors in Eq.~(\ref{eq:FRAG}).}
\label{TABLE0}
\end{table}

\end{document}